\documentstyle[10pt,aaspp4,flushrt]{article}
\def\etal{\it et al. \rm }
\begin{document}

\title{Gas Rich Dwarfs from the PSS-II\\
III. H\,I Profiles and Dynamical Masses}

\author{Jo Ann Eder} \affil{Arecibo Observatory, HC3 Box53995, Arecibo, PR
00612; eder@naic.edu\altaffilmark{1}}

\centerline{and}

\author{James M. Schombert}
\affil{Department of Physics, University of Oregon, Eugene, OR 97403;
js@abyss.uoregon.edu}

\altaffiltext{1}{The Arecibo Observatory is part of the National Astronomy
and Ionosphere Center which is operated by Cornell University under
contract with the National Science Foundation.}

\begin{abstract}

We present Arecibo neutral hydrogen data on a sample of optically selected
dwarf galaxies.  The sample ranges in H\,I mass
from $10^6 M_{\sun}$ to $5\times10^9 M_{\sun}$, with a mean of
$7.9\times10^8 M_{\sun}$.  Using estimated H\,I radii, the H\,I surface
densities range from  0.6 to 20 $M_{\sun}$ pc$^{-2}$, all well below the
critical threshold for star formation (Kennicutt 1998).  $M_{HI}/L$ values
of the LSB dwarfs range from 0.3 to 12 with a mean value of 2.0.

Dynamical masses, calculated from the H\,I profile widths, range from
$10^8 M_{\sun}$ to $10^{11} M_{\sun}$.  There is a strong correlation
between optical luminosity and dynamical mass for LSB dwarfs implying that
the dark matter (whether baryonic or non-baryonic) follows the detectable baryonic
matter.

\end{abstract}

\section{INTRODUCTION}

The study of low surface brightness (LSB) galaxies presents an enormous
observational challenge. LSB galaxies often too faint for more than
limited optical study by broadband imaging.  Spectroscopy and near-IR are
practically impossible due to the fact that too few galaxy photons are
emitted per square arcsec compared to the brightness of the night sky.
Even a nominal number of HST orbits requires rebinning to obtain reliable S/N,
which then loses the high resolution advantage of space imaging.  However,
one intrinsic characteristic has provided an avenue to study this class of
galaxies, a majority of LSB galaxies contain copious amounts of neutral
hydrogen (Schombert \etal 1992, Impey \etal 1996).  This provides both a
quick method to confirm the reality of objects, barely visible in
the optical, while simultaneously providing their distance and crude
kinematic information.

Dwarf galaxies are the most numerous class of galaxies in the Hubble
sequence forming a large component of the number density of galaxies.  For
example, over 80\% of the Local Group are classed as dwarfs and their
space densities are higher even in cluster environments.  They are found
in all types of environments, from dense clusters to the field.  They span
the range of morphological types, from elliptical to highly irregular.
And they cover a full range of current star formation rates, from
quiescent dE's to intense BCD's (Sandage and Binggeli 1984).  Although
dwarf galaxies are low in metallicity (Skillman and Kennicutt 1993), they
are by no means zero-metallicity protogalaxies.  And, while dwarf galaxies are
interesting for their internal properties and star formation histories,
they also serve as probes to the characteristics of dark matter (Ashman
1992) and are tracers of large scale structure (Eder \etal 1989, Zwaan
\etal 1995).

Despite numerous studies on cluster and field dwarfs, it is still unclear
whether dwarf galaxies are a continuation of the spiral sequence or
whether they form a distinct population, possibly fossil remnants from the
era of galaxy formation.  Information about the gas content of dwarf
galaxies is therefore critical to an understanding of their place in the
hierarchy of galaxies.  Unfortunately, our knowledge of the properties of
dwarf irregulars is hampered by small catalogs because of difficulty in
detecting them due to their LSB nature and small angular size.  In order
to improve the sample of dwarf irregular galaxies available for study, we
have identified a new sample of 458 candidate dwarfs from inspection of 49
fields of the new Palomar Sky Survey (PSS-II) directed towards the North
and South Galactic Caps.  Previous papers have presented the catalog and
the optical data (Eder \etal 1989, Schombert, Pildis, and Eder 1997,
Pildis, Schombert and Eder 1997).  This paper presents the single-dish
Arecibo 21 cm data for this sample and determines the dynamical mass
for dwarfs with optical measurements.  The impact of these parameters on
the star formation history of LSB dwarfs is explored in a latter paper
(Schombert, McGaugh and Eder 2000).

\section{OBSERVATIONS}

In this paper, we report on the 21 cm observations of three samples of
dwarf galaxy candidates selected visually from photographic plates of the
Second Palomar Sky Survey (PSS-II).  All of the samples were chosen so as
to optimize their observation with the Arecibo radio telescope, meaning
they lie between $-5$ and $+35$ in declination.  The first sample (Sample
1) of 178 candidates was obtained from 15 fields located on regions of the
fall sky occupied by a nearby void.  The plates used were high quality
preproduction plates.  Of these candidates, 132 were observed at 21 cm.
The void is centered at RA = 0h45m, Dec = +20, v = 3500 km sec$^{-1}$, and
is of roughly spherical shape with a diameter of 1500 km sec$^{-1}$.  The
second sample (Sample 2) of 278 was identified using methods similar to
those for the first sample from 35 random fields between RA's of 7h to 16h
and Dec's of 10 to 25, irrespective of galaxy density.  Of these, all were
observed at 21 cm.  The third much smaller sample of 27 (Sample 3)
consisted of new dwarf candidates identified from a visual reexamination
of one of the fields of Sample 1 (Field 409), using the final PSS-II
production plates.  Sample 3 also included observations of some of the
dwarf candidates from Sample 1 which were not originally observed with 21
cm due to a lack of telescope time.

All objects on the plates with angular diameters greater than 20 arcsecs, 
irregular appearance (Sm, Im, Irr, dI) and low surface brightness
were selected.   Low mass implies instability, at least with respect to
the pattern of star formation.  Thus, irregular structure is nominally
associated with dwarf galaxies (unless the star formation is completed,
as in a dwarf elliptical).  Basically, late-type galaxies can be broken
down into three morphological types; those with bulges and LSB disks
(Sc and Sd class, those with apparent axial symmetry but no bulges (Sm
class) and those with no symmetry (Im class).  However, morphology is
only weakly related to the mass of a galaxy.  There are many irregular
galaxies that are massive (Hunter and Gallagher 1986) and many spirals
which are dwarf-like in luminosity and size (Schombert \etal 1995).  In
addition, many high luminosity events, such as galaxy-galaxy
interactions, are also associated with irregular morphology, although
usually of a nature that identifies them as tidal in origin.  Irregular
morphology, by itself, is not a sufficient criteria for dwarf
classification, but is sufficient for the development of a candidate
list.  The coordinates of each candidate dwarf were taken directly from
the plate material using a fine ruler and the position of SAO
stars.  Accuracy of the coordinates were typically $\pm$15 arcsecs based
on the centering of CCD imaging, which is sufficient for detection with
the Arecibo beam of 3 arcmin diameter.  Further details of the
selection of the three samples can be found in our previous papers:
Eder \etal 1989 (Sample 1) and Schombert, Pildis, and Eder 1997 (Samples 2
and 3).

Follow-up CCD images in V and I for most of those galaxies detected in H I
from Sample 2 were obtained on the Hiltner 2.4m telescope located at
Michigan-Dartmouth-M.I.T. (MDM) Observatory.  We obtained images using
either a Thomson 400$\times$576 pixel CCD (0.25 arcsec pixel$^{-1}$) or a
Ford-Loral 2048$\times$2048 pixel CCD binned 3$\times$3 (0.51 arcsec
pixel$^{- 1}$), with minimal exposure times of 25 minutes in Johnson $V$
and 15 minutes in Johnson $I$.  Analysis of these images reveal that the
candidates are gas-rich dwarf galaxies with total luminosities of $M_V >
-18$ mags, central surface brightnesses between 21 and 23 I mags
arcsec$^{-2}$, exponential scale lengths of typically less than 3 kpc and
mean isophotal diameters (of approximately the Holmberg diameter) of 8 kpc
(as compared with a mean of 30 kpc for the UGC galaxies, Pildis, Schombert
and Eder 1997).  A sample of comparable size to ours would be the Virgo
Cluster dwarfs (Hoffman \etal 1997) but these dwarfs reside in a cluster
environment while ours inhabit a range of environmental densities.

We observed the dwarf candidates for the H\,I line at 21 cm with the
Arecibo 305m telescope during the 1988, 1992 and 1993 observing season.
All observations were made with the 21 cm dual-circular feed positioned to
provide a maximum gain (8 K Jy$^{-1}$) at 1400 MHz.  The 2048 channel
autocorrelator was used and the independent, opposite polarized signals
were each divided into two subcorrelators of 512 channels.  In order to
search a larger velocity space, the secondary local oscillators of each
polarization set of subcorrelators were offset on either side of the
standard local oscillator frequency of 260 MHz by 8.75 MHz, allowing a
total velocity coverage of 8000 km sec$^{-1}$, a velocity resolution of 8.6
km sec$^{-1}$, and some overlap at the band edges.  The observations were
centered on 4000 km sec$^{-1}$, which avoided detection of the strong
Galactic hydrogen signal on the low-velocity end, and extended to 8120 km
sec$^{-1}$.  Observations were made in the total power mode with 5 minute
ON- source and OFF-source observations.  In most cases, only one 5-minute
ON-source integration was required for detection.  Wherever possible, the
zenith angle was kept less than 14 degrees to minimize the degradation of
the gain.

Efforts were made to assure that the signal was due to the candidate
galaxy. The field of each position was examined on the original plates and
also with NED to check for possible sources of confusion.  Because many of
the detections were very narrow, the possibility existed that they were
actually caused by radio frequency interference (RFI).  However, RFI
confusion can be distinguished by their polarization properties and/or,
with 3-level sampling, by the ringing they cause in the band which is
eliminated with hanning smoothing.  Those narrow detections that were
still suspicious were reobserved to confirm their authenticity.  Higher
resolution observations of several narrow line detections revealed
double-horned features.

\section{ANALYSIS}
\subsection{Galaxy Data}

All three samples were analyzed and reduced in the same fashion.
Calibration of the spectra for atmospheric correction and noise diode
variation with frequency was performed during the phase of translation
into export format for compatibility with the Arecibo Observatory
ANALYZ-GALPAC analysis system.  Baseline fitting, removal, and parameter
extraction were all performed using the latter package.  The width,
referred to hereafter as $W_{50}$, is defined as the full width across the
profile measured at a level of 50\% of the peak flux.

In Table 1 we present the results of the analysis for the 250 galaxies
whose digital spectra have been examined.  In addition, 143 galaxies which
were searched for, but not detected, are also listed in Table 1.  The
nominal search range was 1000 to 8,000 km sec$^{-1}$, with occasional
extensions to 12,000 km sec$^{-1}$.  The typical RMS values were 3 mJy.
All distance related values in this paper use values of $H_o = 75$ km
sec$^{-1}$ Mpc$^{-1}$.  All the data included in this paper can also be
found in ASCII format at zebu.uoregon.edu/$\sim$js.

Details of the entries in Table 1 are as follows:

\begin{itemize}

\item Column (1): Object ID, based on the PSS-II field number and a running
number for the dwarfs found on that field.

\item Columns (2) and (3): Right ascension and declination in the 1950.0
epoch.  These are measured off the PSS-II plates themselves and are
accurate to 15 arcsecs.

\item Column (4): The heliocentric
velocity, $V_h$ of the H\,I line, taken at
the midpoint of the profile (50\% of peak level).

\item Column (5): The velocity corrected for the motion of the Sun
relative to the center of mass of the Local Group, according to the
precepts of de Vaucouleurs, de Vaucouleurs, and Corwin (1976), $V_o$,
in km sec$^{-1}$.

\item Column (6): The velocity width, $W_{50}$, in km sec$^{-1}$.  No
corrections have been made for inclination or turbulent broadening.

\item Column (7): The profile shape where S means single-horned and D means 
double-horned. A semi-colon indicates uncertain classification.

\item Column (8): The observed integrated 21-cm H\,I line flux, $S =
\int S_V dV$ in Jy km sec$^{-1}$.

\item Column (9): The H\,I gas mass of the object, in $M_{\sun}$,
calculated using the standard conversion from flux to gas mass.

\item Column (10): Observing run number from which the data was taken,
where run \#1 is Sample 1 observed in 1986, \#2 is Sample 2 and \#3 is
Sample 3, both observed   in 1992 and 1993.

\end{itemize}

\subsection{Notes on Individual Galaxies}

{\it 409-4,476-5} -- Velocity scale in the spectrum is not correct because
the LO synthesizer was set incorrectly.  The velocities in the Table are
the correct values.

{\it 409-7} -- 1.5 arcmin from SRGb 062.12 (v=7333 km sec$^{-1}$)

{\it 409-11} -- 8.5 arcmin from 409-10

{\it 409-16} -- 6.1 arcmin from N7773 (v=8486 km sec$^{-1}$)

{\it 409-17} -- 3.0 arcmin from KUG 2356+303

{\it 409-18} -- 1.6 arcmin from U152 (v=4863 km sec$^{-1}$)

{\it 474-9} -- 8.1 arcmin from N304 (v=4991 km sec$^{-1}$

{\it 476-2} -- also U948

{\it 476-4} -- also U1073

{\it 476-5} -- also U1084

{\it 495-1} -- 5.8 arcmin from U4361 (v=3741 km sec$^{-1}$)

{\it 495-3} -- detected as F495-V1 (LSB2)

{\it 508-1} -- The dwarf HI signal could be embedded in that of NGC 4961,
detected here off center, but is buried in flux from primary galaxy.  NGC 4961
detected by Shostak.  The velocities and widths listed in the Table are
measured at 50\% of the mean flux

{\it 508-2} -- near A1656

{\it 512-1} -- lots of RFI

{\it 512-4} -- 8.8 arcmin from KUG 1430+264

{\it 512-5} -- observed in velocity range 0 - 12,000 km sec$^{-1}$

{\it 512-10} -- 6.5 arcmin from IC4475

{\it 514-2} -- observed in velocity range 0 - 12,000 km sec$^{-1}$

{\it 514-3} -- paired with 514-4

{\it 514-5} -- The spectrum is missing from the Figure

{\it 514-6} -- observed in velocity range 0 - 12,000 km sec$^{-1}$

{\it 516-1} -- observed in velocity range 0 - 12,000 km sec$^{-1}$

{\it 538-1} -- 4.9 arcmin from U3 (v=7882 km sec$^{-1}$)

{\it 538-2} -- 7.8 arcmin from N7817 (v=2309 km sec$^{-1}$)

{\it 538-15} -- also U12843

{\it 538-16} -- also U12846

{\it 538-18} -- 8.3 arcmin from U12916 (v=6352 km sec$^{-1}$)

{\it 539-2} -- also U159

{\it 539-5} -- also IC1542

{\it 539-6} -- also U200

{\it 540-2} -- 7.9 near U425 (v=5841 km sec$^{-1}$)

{\it 540-3} -- also U477

{\it 540-8} -- 1.3 arcmin from F474-10

{\it 559-3} -- Malin object, 9.3 arcmin from U3920

{\it 563-2} -- detected as F563-V2 (LSB2)

{\it 563-3} -- 7.7 arcmin from U4588 (v=4264 km sec$^{-1}$), detected as F563-V1 (LSB2)

{\it 563-4} -- dwarf spiral, detected as F563-1 (LSB1)

{\it 564-2} -- measured parameters are uncertain because of overlapping RFI 

{\it 564-8} -- detected as F564-V3 (LSB2)

{\it 564-12} -- 7.4 arcmin from U4729 (v=3900 km sec$^{-1}$)

{\it 564-15} -- dwarf spiral

{\it 565-2} -- UGC 5005, detected by Bicay and Giovanelli

{\it 565-3} -- possibly two merged signals - several nearby galaxies are
within 1 arcmin on the CCD field

{\it 565-5} -- signal is blend of UGC 5086 and NGC 2903 in the beam and
sidelobes

{\it 565-10} -- KARA 68-056, not detected as F565-V4 (LSB2)

{\it 568-3} -- REIZ 363, 3.4 arcmin from U5629 (v=1236 km sec$^{-1}$)

{\it 568-4} -- ZWG 094.005

{\it 568-5} -- observed in velocity range 0 - 12,000 km sec$^{-1}$ but had
strong RFI (rms=2.3mJy)

{\it 570-3} -- detected as F570-V1 (LSB2)

{\it 570-4} -- detected as F570-7 (LSB2), 4.5 arcmin from IC2703

{\it 570-6} -- poor S/N, companion to D571-5

{\it 571-3} -- 8.7 arcmin from AGC210538 (v=6196 km sec$^{-1}$)

{\it 571-5} -- two galaxy signals appear in the spectrum, due to two
galaxies within 20 arcsec on the CCD image.  The parameters listed in the
table are from the face-on diffuse dwarf galaxy.  The other signal is
double peaked and therefore emanates from the inclined spiral ($v$=5985 km
sec$^{-1}$, W=86 km sec$^{-1}$, FI = 157.2 mJy km sec$^{-1}$ )

{\it 571-6} -- also not detected as F571-10 (LSB2)

{\it 572-5} -- 8.3 arcmin from U6770

{\it 575-1} -- I3840, detected as F575-2 (LSB2)

{\it 575-2} -- UGC 8011

{\it 575-3} -- I4107

{\it 575-4} -- pair to 575-3

{\it 575-5} -- KARA 68.215, in same beam as F575-3 detected in LSB2. The
velocities and widths listed in the Table are measured at 50\% of the mean
flux

{\it 576-2} -- strong continuum

{\it 576-6} -- ZWG 101.005

{\it 576-7} -- measured parameters uncertain because of interference

{\it 576-11} -- MCG +03-34-034, 2.7 arcmin from U8448 (v=7151 km sec$^{-1}$),
probably U8448 off band

{\it 577-5} -- dwarf spiral

{\it 577-6} -- measured parameters uncertain because the galaxy spectrum
is near the edge of the band, companion to D648-4

{\it 582-1} -- probably off-beam detection of F582-2 (LSB2) 

{\it 582-4} -- 9.5 arcmin from F583-2

{\it 584-2} -- also U10140, 6.1 arcmin from U411A (v=2525 km sec$^{-1}$)

{\it 609-2} -- also U385

{\it 609-3} -- companion to D609-2

{\it 609-4} -- UGC 472, detected by Bothun \etal 1985

{\it 609-5} -- also U560

{\it 611-5} -- detected as F611-1 (LSB2)

{\it 611-6} -- also U883

{\it 611-7} -- also U891

{\it 611-11} -- 6.1 arcmin from F611-V2

{\it 611-13} -- also U1026

{\it 611-14} -- also U1056

{\it 611-18} -- 8.7 arcmin from U1087 (v=4485 km sec$^{-1}$)

{\it 631-3} -- questionable signal at v= 168 km sec$^{-1}$, W = 33, FI =
123 mJy km sec$^{-1}$

{\it 631-4} -- questionable signal at v= 352 km sec$^{-1}$, W = 50, FI =
147 mJy km sec$^{-1}$  631-7 UGC 4115, detected by Bicay and Giovanelli

{\it 631-7} -- also U4115

{\it 631-8} -- dwarf spiral

{\it 631-9} -- questionable signal at v= 2280 km sec$^{-1}$, W = 59, FI =
178 mJy km sec$^{-1}$

{\it 631-11} -- questionable signal at v= 6130 km sec$^{-1}$, W = 60, FI =
123 mJy km sec$^{-1}$

{\it 637-1} -- very noisy spectrum

{\it 637-5} -- 7.8 arcmin from IC581

{\it 637-7} -- 7.5 arcmin from U5342 (v=4560 km sec$^{-1}$) (pair?)

{\it 637-10} -- observed in velocity range 0 - 12,000 km sec$^{-1}$

{\it 637-21} -- observed in velocity range 0 - 12,000 km sec$^{-1}$

{\it 640-2} -- 640-2,3 and 4 are all in the same beam (UGC 6006, UGC
6007).  Signal is due to U6006 also detected by Schneider \etal

{\it 640-7} -- UGC 5948, detected by Schneider \etal

{\it 640-8} -- near NGC 3489 

{\it 640-10} -- UGC 5944

{\it 640-11} -- measured parameters uncertain due to overlapping RFI, 5.4 arcmin
from N3412 (v=865 km sec$^{-1}$)

{\it 640-14} -- near ZWG 066.072 

{\it 640-15} -- NGC 3559, detected by Bicay and Giovanelli

{\it 646-4} -- Rea 66 

{\it 646-5} -- 646-6 is also in the beam, signal could be from either one

{\it 646-7} -- UGC 8091, detected by Lewis, Helou, and Salpeter

{\it 646-11} -- UGC 8061 - detected by Schneider \etal

{\it 648-1} -- solar interference

{\it 648-2} -- very bad baseline and noise

{\it 648-4} -- companion to D577-6

{\it 648-6} -- solar interference

{\it 651-6} -- comp to ZWG 076.018

{\it 656-1} -- U10398 - detected by Schneider \etal

{\it 656-2} -- U10281 - detected by Schneider \etal

{\it 656-4} -- observed in velocity range 0 - 12,000 km sec$^{-1}$

{\it 656-5} -- 9.2 arcmin from U10218 (v=1080 km sec$^{-1}$)

{\it 685-6} -- also U1323

{\it 685-18} -- also U1515

{\it 685-19} -- remeasured parameters since Eder \etal, 4.4 arcmin from IC1772 

{\it 709-5} -- KARA 68.060

{\it 709-7} -- 9.5 arcmin from N3130 (v=8206 km sec$^{-1}$)

{\it 709-9} -- 9.2 arcmin from U5304 (v=12308 km sec$^{-1}$)

{\it 709-10} -- 5.3 arcmin from U5304 (v=12308 km sec$^{-1}$)

{\it 721-1} -- ZWG074.037 also in beam

{\it 721-5} -- dwarf spiral, detected as F721-V4 (LSB2)

{\it 721-7} -- 4.1 arcmin from N5463 (v=7235 km sec$^{-1}$) (pair)

{\it 721-8} -- companion to D721-9

{\it 721-10} -- limits of the measured galaxy spectrum are uncertain

{\it 721-14} -- signal is from U8995 at 140219+090224 detected by Bothun \etal 1985

{\it 721-15} -- broad wings - parameters uncertain - NGC 5511 interaction

{\it 723-3} -- signal on edge of spectrum, parameters uncertain

{\it 723-4} -- 6.3 arcmin from U9356 (v=2225 km sec$^{-1}$)

{\it 723-9} -- also U9500

{\it 749-3} -- also U12416

{\it 749-6} -- also U12480

{\it 749-7} -- 5.9 arcmin from N7587 (v=8917 km sec$^{-1}$)

{\it 749-8} -- 5.0 arcmin from N7593 (v=4108 km sec$^{-1}$)

{\it 749-9} -- 5.8 arcmin from U12497 (v=3761 km sec$^{-1}$)

{\it 749-10} -- also U12497

{\it 749-12} -- also U12553

{\it 749-14} -- also U12562, 8.2 arcmin from N7641 (v=7872 km sec$^{-1}$)

{\it 749-17} -- 8.4 arcmin from U12687 (v=6158 km sec$^{-1}$)

{\it 774-1} -- dwarf spiral

{\it 774-4} -- PK 215+11.1

{\it 822-4} -- remeasured parameters since Eder \etal, 3.7 arcmin from D822-5

{\it 822-6} -- also F750-2

{\it 822-14} -- 8.8 arcmin from N7757 (v=2956 km sec$^{-1}$) (pair)

\subsection{Mass Estimates}

To determine the dynamical masses from the H\,I profiles, inclinations and
metric sizes are needed.  To this end, CCD images were obtained for 116
galaxies at the 2.4m MDM telescope.  The isophotometric ellipse fitting
was done as described in Pildis, Schombert and Eder 1997.  Scale lengths
were derived from least-squares fitting of an exponential function to the
data in the surface brightness-radius plane.

Due to their LSB nature, the inclination measurements of dwarf galaxies
are problematic (Tully \etal 1978, Lo \etal 1993).  Since H\,I rotation
velocities are being used, strictly speaking H\,I diameters should also be
used.  However, numerous studies have shown that the H\,I surface
densities follow the optical surface densities (see McGaugh 1992).  Thus,
we adopt the optical parameters for inclination as based on the
eccentricity of the ellipse at the 25 mag arcsec$^{-2}$ isophote.  The
$R_{25}$ isophotal radius, axial ratio and $I$ magnitude are listed in
Table 2.

The dynamical mass is determined from the virial theorem using the
prescription of Staveley-Smith, Davies and Kinman (1992, hereafter SDK).
However, to complicate matters, ellipticity studies indicate that dwarf
galaxies are primarily triaxal with particular signatures in their
structural parameters that confirm their non-oblate nature (Schombert,
McGaugh and Eder 2000).  If dwarf galaxies are not oblate rotators, then
as discussed in SDK, the use of a pure rotation term to calculate the
dynamical mass will underestimate this value by as much as 30\%.  Hence,
we adopt SDK's virial mass estimate, $M_{dyn}$, as given by

$$ M_{dyn} = 2.3\times10^5 (V^2_{rot} + 3\sigma^2) R_{HI} $$

\noindent where $V_{rot}$ is the rotation velocity, $\sigma$ is the
isotropic velocity dispersion and $R_{HI}$ is the outer radius of the H\,I
emission.  The velocity dispersion is assumed to be 10 km sec$^{-1}$, the
standard value for late-type galaxies (Shostak and van der Kruit 1984).
High resolution H\,I maps of gas-rich dwarf galaxies also indicate a
typical velocity dispersion of 10 km sec$^{-1}$ (Skillman \etal 1988, Lo
\etal 1993).  Note that, for dwarfs without rotation, $V_{rot}$
could be replaced with $\sigma$ and the dynamical masses is underestimated
by a factor of 3, but we have assumed rotation in our discussion.

We calculate $V_{rot}$ from the measured H\,I profile width,
correcting for inclination and velocity dispersion such that

$$ W_{50} = 2(V_{rot} {\rm sin}\, i + 1.18\sigma)$$

\noindent where the inclination, $i$, is determined from the axial ratio
of the outer isophotes of the dwarf galaxy.  The remaining variable is
$R_{HI}$, the radius of the edge of the H\,I emission.  We can measure
$R_{25}$, the radius of the 25 $B$ mag arcsec$^{-2}$ isophote directly
from the optical images.  H\,I mapping studies indicate that the H\,I
diameter of dwarf galaxies is much larger than the optical radius.  Van
Zee, Skillman and Salzer (1998) found that the H\,I diameter is
approximately 4 times the optical radius for a mapping study of 5 dwarfs.
However, these dwarfs were selected as having an overabundance of H\,I.  
Salpeter and Hoffman (1996) discovered a correlation between optical and
H\,I radius, such that $R_{HI} = 2.3 R_{opt}^{-0.11}$ for a sample of 109
dwarf irregular galaxies.  Based on this evidence, we have set $R_{HI} =
2.3 R_{25}$ for the dwarf sample and determined all scale length values
based on this conversion.  The calculated values $M_{dyn}$ are listed in
Table 2 along with the ratio of $M_{H\,I}/M_{dyn}$.

\section{DISCUSSION}

\subsection{What is a dwarf galaxy?}

Our understanding of dwarf galaxies, their properties, contents and star
formation histories, has changed sharply, mostly because our catalogs of
dwarf galaxies have expanded in the last decade.  Our initial view of dwarf
galaxies were based on our studies of the LMC, SMC and other Irr class 
galaxies in the Local Group (Holmberg 1957) and low luminosity dwarf
spheriodals such as Leo A.  These objects, although interesting in their
own right, do not properly represent the range in mass, appearance and
star formation history that define what we consider today to be the class
of dwarf galaxies.  For example, in the late 1950's, de Vaucouleurs (1959)
extended the late-type Hubble classes into the Sd, Sm and Im types to
delineate the gradual loss of spiral structure.  Most of these extreme
late-type galaxies would later be defined as dwarfs based on their optical
magnitudes (see Sandage and Binggeli 1984); however, they do not define
the dwarf galaxy population.

Inspection of nearby clusters (Virgo and Fornax) revealed an extensive
population of gas-poor elliptical dwarfs and gas-rich irregular dwarfs,
subsequently called dE and dI to indicate their morphologically link
to ordinary ellipticals and irregulars.  The similarities between the dwarf
ellipticals and irregulars to their giant counterparts (such as
mass-metallicity relations, fundamental plane parameters and star
formation) did appear to simplify the Hubble sequence as one divided by
morphological class on one axis and by luminosity (and therefore
mass) on the other.  However, later work on field dwarfs (Eder \etal 1989,
Schombert \etal 1997) showed that there is very little difference between
the cluster dI class and the de Vaucouleurs Im class.

With the addition of a wide variety of morphological types, the definition
of a dwarf galaxy has become blurred in the last decade.  If asked, a
theorist would define a dwarf galaxy by mass, then consider various
formation scenarios that produce a mass spectrum.  Unfortunately, dark
matter becomes fractionally more significant in galaxies with decreasing
mass (Ashman 1992), and this means that galaxy mass does not map, in a
direct fashion, into observables such as luminosity, size or even
rotational velocity.  In addition, to make a difficult situation even
worse, lower luminosity, density and size make the necessary observables
laborious or impossible to obtain with our current technology and/or
limited telescope time.

Within the observables, there are several methods in which to define a
dwarf galaxy.  For example, one can define a dwarf galaxy by optical
luminosity (Tammann 1980).  Since a galaxy's optical luminosity is the
primarily due to the photons emitted by stellar photospheres, then the
total luminosity should reflect the stellar mass.  With a stable IMF, the
connection between luminosity and mass is parameterized by $\Upsilon_* =
M_*/L$, the stellar mass, $M_*$ divided by the total luminosity of a
galaxy, $L$.  Therefore, a dwarf galaxy could be defined by some boundary
in optical luminosity as it reflects into stellar mass.  However, since
the fraction of dark matter (baryonic and non-baryonic) increases with
later type galaxies (Ashman 1992), the use of luminosity to trace mass
begins to decouple or has too many variables.

Dwarf galaxies are typically low in density, so division by central
surface brightness (based on fits to the galaxy's profile) can be used as
an indicator of dwarfness.  There is some physical reasoning to support
the idea that all LSB galaxies should be dwarf systems.  To produce low
stellar densities would require very low past star formation rates and,
hence, low total stellar mass and luminosity.  Low surface brightness does
not necessarily mean low stellar densities, but the colors of LSB galaxies
rule out any highly unusual $\Upsilon_*$ (McGaugh and de Blok 1997)
implying that low optical density is indeed low stellar density.   Or,
conversely, we might expect a dwarf galaxy of low mass to have low mean
density which, in turn, inhibits star formation to produce a LSB object.
One could argue that initial conditions, such as the amplitude of the
density perturbation in the early Universe from which the galaxy will
form, produces a necessary connection between a dwarf galaxy and a LSB
galaxy.  Unfortunately, the discovery of large Malin galaxies destroyed
the notion that all LSB galaxies are dwarfs. In fact, most LSB disk
systems are comparable in size to the ordinary spirals that make up the
Hubble sequence (McGaugh, Schombert and Bothun 1995).

Low mass implies instability, at least with respect to the pattern of star
formation.  So, irregular structure is nominally associated with dwarf
galaxies (unless the star formation is completed, as in a dwarf
elliptical).  Basically, late-type galaxies can be broken down into three
morphological types; those with bulges and LSB disks (Sc and Sd class,
those with apparent axial symmetry but no bulges (Sm class) and those with
no symmetry (Im class).  However, morphology is only weakly related to the
mass of a galaxy.  There are many irregular galaxies that are massive
(Hunter and Gallagher 1986) and many spirals which are dwarf-like in
luminosity and size (Schombert \etal 1995).  In addition, many high
luminosity events, such as galaxy-galaxy interactions, are also associated
with irregular morphology, although usually of a nature that identifies
them as tidal in origin.  Irregular morphology, by itself, is not a
sufficient criteria for dwarf classification.

The most promising approach is to use the size of a galaxy as a measure of
its dwarfness.  A combination of angular size and irregular morphology
were the criteria used in the initial construction of the PSS-II LSB dwarf
catalog.  Our previous LSB surveys (Schombert and Bothun 1988) was rich in
LSB disks with scale lengths greater than 4 kpc, but few dwarf galaxies.
That survey used a minimal angular size of one arcmin, the same as the
UGC, to determine the bias against LSB galaxies in our catalogs.  This
PSS-II LSB dwarf survey set the criteria to be between 30 arcsecs and one
arcmin, with a dramatically different metric size distribution as seen in
Figure 3 of Schombert \etal (1997).  With the acquisition of surface
photometry and a H\,I redshift, a proper comparison of scale length sizes
for the sample can be made (shown in the top panel of Figure 3).  Most of
the objects (90\% of the sample) turn out to be quite small, less that the
typical disk galaxy (3.2 kpc, de Jong 1996), testifying to the power of
using morphology to select dwarf galaxies.  We suspect that many of the
undetected dwarf-like objects are probably background Sc's outside the
velocity range of the observing set-up for the Arecibo telescope.  We
conclude from Figure 3 that a robust method of collecting a sample of
dwarf galaxies is to select objects by irregular morphology over a limited
angular size (in our case, 30 to 60 arcsecs).

\subsection{H\,I Profiles}

A sample of 458 candidates were found by visual inspection of the PSS-II
plates.  Each candidate object was assigned a quality index as a measure
of its probability to be a dwarf galaxy based on its morphological
appearance.  This is a purely subjective system, based on an objects
irregular shape, lack of coherent spiral patterns, small to non-existent
H\,II regions and a general LSB appearance.  Most of these galaxies would
be classified as extreme late-type (Sm, Im, Irr or dI) and a more complete
discussion of their morphological and properties can be found in
Schombert, McGaugh and Eder (2000).  From our initial sample, 432 with
high quality indices were observed on the Arecibo 300m telescope at 21-cm
out to a velocity of 8,000 km sec$^{-1}$.

Of the 432 galaxies observed, 259 (60\%) were detected.  The objects not
detected are probably background LSB spirals (with spiral patterns that
were not visible on the plates), gas-poor dwarfs (dE's) or dwarfs whose
H\,I masses and distance fall below the sensitive limit of the telescope.
A small sample of the nondetections were reobserved in a velocity range of
4,000 to 12,000 km sec$^{-1}$ in order to determine how many of the nondetected
candidates were actually distant spiral galaxies.  Of 18 candidates, 10
were detected at the higher velocities.  However, only 4 of these have
velocity widths which suggest a normal spiral galaxy (one being the Malin
object D559-3 with a velocity width of 576 km sec$^{-1}$).  The loss of
gas-poor dwarfs from the sample is unfortunate since the distribution of
dE's, outside of a cluster environment, is not known.  In fact, the
existence of a true dE, independent of a high local density or nearby
companion, has not been confirmed (Binggeli, Tarenghi and Sandage 1990).

The spectra of all the detections from the three samples are presented in
Figure 1.  A hanning smoothing function has been applied, followed by a
boxcar of 3.  Each panel displays the raw data as well as the fitted
baselines.  A majority of the sample requires only a 2nd or 3rd order
polynomial to flatten the baseline.  As seen by other studies, most dwarfs
have the classic double-horned profile indicative of rotation, although
the data is too low in resolution to discern features such as asymmetries,
filled-in cores or extended line wings (Hoffman \etal 1996, Matthews, van
Driel and Gallagher 1998, Staveley-Smith, Davies and Kinman 1992).  Skewed
profiles may be due to confusion with nearby sources or pointing errors.

The H\,I profiles can be divided into two types, gaussian and
double-horned.  A galaxy with a flat or rising rotation curve, and a
distribution of gas that declines with radius in the form of a power law,
will exhibit a H\,I profile that has two peaks with a flat plateau region
between them (see Giovanelli and Haynes 1988).  The 246 spectra in Figure
1 were classified as either double horned, single horned or uncertain
(listed in Table 1).  Double horned profiles (such as D475-4 or D704-2)
have distinguishable peaks or a plateau shape.  Single horned profiles
display a smooth gaussian shape with no indication of flattening.  In many
cases (e.g.  D709-5), the profile is broad, but the double horn shape is
unclear due to noise.  In other cases (e.g. D702-1), there is an asymmetry
that distorts the profile.  Profiles with noisy plateaus are classed as
uncertain.  For the 246 spectra, 140 are double horned, 61 are single
horned and 45 are uncertain.  Assuming the uncertain profiles are evenly
divided between single and double (most are broad and probably double
horned), then double horned profiles are found in at least 70\% of the
sample.

The distribution of H\,I fluxes and H\,I profile widths are shown in
Figure 2.  All the detections are well above the 3$\sigma$ level given the
mean fluxes and velocity widths.  The mean velocity width, even without
inclination corrections, is low compared to a typical late-type galaxy
sample (see Schombert, Pildis and Eder 1997) confirming our intent to
acquire a sample of low mass objects.  A handful of objects with $W_{50} >
150$ km sec$^{-1}$ represent a few Malin type objects, irregular in their
morphology, that mimic a dwarf galaxy's appearance yet are actually large,
distant spirals.

The metric scale length ($\alpha$) distribution, discussed in \S4.1, is
shown in the top panel of Figure 3.  Based on exponential fits to the $I$
band surface brightness profiles, $\alpha$ provides the most fundamental
property of a galaxy, its characteristic size.  While the isophotal size
of a galaxy, combined with its mean surface brightness, determines its
visibility and detection probability.  The typical scale length for a disk
galaxy is indicated.

The redshift distribution of the detected dwarfs is shown in the bottom
panel of Figure 3.  The spatial distribution was shown in Figure 6 of
Schombert, Pildis and Eder (1997).  To repeat that paper's conclusion, LSB
dwarfs trace the same large scale structure as brighter galaxies.  Figure
3 confirms that a majority of the dwarfs are located between 2000 and 6000
km sec$^{-1}$, although there is a strong mass dependence.  This dependence
on mass is shown in the lower panel of the Figure 3 as a plot of H\,I mass
versus redshift.  The solid line shows the telescope 3$\sigma$ detection
limit for a dwarf galaxy with a line width ($W_{50}$) of 75 km sec$^{-1}$.
A majority of the detections lie above this line.  Objects below the limit
all have line widths smaller than the assumed 75 km sec$^{-1}$ and, thus, a
higher S/N per channel.  Note that the lowest mass object
($M_{HI}=2.5\times10^5 M_{\sun}$) is located at less than 500 km
sec$^{-1}$.  This plot demonstrates that the search for very low mass
H\,I-rich galaxies is extremely difficult as a combination of confusion
with Galactic H\,I emission and small angular size makes their inclusion
in any catalog problematic (see Schneider and Schombert 1999).  It is also
apparent from Figure 3 that a search for objects with $M_{HI} < 10^7
M_{\sun}$ will require a smaller angular limit in order to catalog
candidates and deeper H\,I observations to obtain sufficient S/N (see also
Zwaan \etal 1997) and that contrary to expectations (Briggs 1997) our
galaxy catalogs are incomplete even in the local region of the Universe.

\subsection{H\,I Masses and $M_{HI}/L$}

The distribution of H\,I masses, calculated using the standard formula of
distance and H\,I flux (Giovanelli and Haynes 1988) is shown in the top
panel of Figure 4.  The sample has a mean value of $M_{HI}=7.9\times10^8
M_{\sun}$ and a sharp cutoff at $M_{HI}=4.0\times10^9 M_{\sun}$.  This can
be compared to the UGC H\,I sample (see Figure 4, Schombert, Pildis and
Eder 1997) where the mean H\,I mass of a UGC selected galaxy is
$M_{HI}=3.0\times10^9 M_{\sun}$.  Figure 4 shows that, whereas LSB
galaxies are gas-rich, this does not mean that LSB dwarfs have high gas
masses.  The distribution of H\,I masses for this sample still defines the
low-end of a gas mass distribution from any galaxy catalog and a galaxy
selected to be dwarf-like by H\,I mass is as valid as classification by
total mass or luminosity.

The bottom panel of Figure 4 displays mean H\,I surface density
($\Sigma_{HI}$) for each dwarf with available optical imaging.  The H\,I
radius has been used to determine the average surface density (assuming
the same axial ratio as the outer optical isophote).  The $\Sigma_{HI}$
values range from 0.6 to 20 $M_{\sun}$ pc$^{-2}$ with a mean value of 5.
The importance of $\Sigma_{HI}$ to star formation was outlined in
Kennicutt (1989) where he found that star formation follows a power-law in
ordinary galaxies (the so-called Schmidt law, SFR $\propto
\Sigma_{HI}^\alpha$).  Key to Kennicutt's findings was that there exists a
threshold for star formation based on the Toomre (1964) instability
criterion, the balance between gravity and rotation/thermal pressures.  In
a study of five dwarf galaxies, van Zee \etal (1997) found that the
critical threshold was between $9\times10^{20}$ and $6\times10^{21}$ atoms
cm$^{-2}$, which corresponds to the range of 5 and 30 $M_{\sun}$ pc$^{-2}$
in H\,I mass after corrections for neutral helium and other metals.  This
indicates that the mean H\,I surface densities for most of the LSB dwarfs
in this sample are below the critical threshold density for star
formation.  Of course, there can exist local density enhancements that
could form sites of star formation. However, the average surface density
for the galaxy is below the threshold and this would explain the LSB
nature to the dwarf sample, where star formation has occurred in the past
(since there is some measurable stellar luminosity), but that this star
formation has been inhibited and the current star formation levels are
very low (McGaugh and de Blok 1997).

The fraction of a galaxy's mass in the form of gas increases with later
Hubble types and can dominate the kinematics of a dwarf galaxy (Meurer
\etal 1996).  In addition, the amount of gas in dwarf galaxies increases
with decreasing surface brightness to the point where many of these dwarfs
have gas fractions between 0.7 and 0.9 whereas the typical spiral have
$f_g \approx 0.4$ (Schombert, McGaugh and Eder 2000).  Thus, the Hubble
sequence can be described as a declining ratio of stellar to gaseous
material, where early-type galaxies are dominated by stars, later types by
gas.  This ratio is typically expressed as $M_{HI}/L$ and is shown in
Figure 5 plotted against galaxy central surface brightness ($\mu_o$) and
galaxy color ($V-I$).  The LSB dwarfs have a much higher value of
$M_{HI}/L$ then the ordinary spirals, taken from the de Jong (1996).   The
distribution of $M_{HI}/L$ herein is similar to that of van Zee, Haynes
and Giovanelli (1995), a study of extreme $M_{HI}/L$ galaxies from the
Haynes and Giovanelli H\,I survey with the $M_{HI}/L$ values ranging from
0.3 to 12 and a mean value of 2.0.  This average $M_{HI}/L$ found here is
the same as the value found by van Zee, Haynes and Giovanelli (after
adjustment for the $I$ band luminosities) and we also find very few
galaxies with $M_{HI}/L > 10$ even though this sample was selected by
optical morphology rather than $M_{HI}/L$ ratio.

By itself, the H\,I properties of LSB dwarfs are poorly correlated with
optical surface brightness or color.  But, when compared with the de Jong
sample, it is clear that the dwarfs continue the trend of increasing
$M_{HI}/L$ with decreasing surface brightness and bluer colors.  Both
these correlations provide insight into the star formation history of
dwarf galaxies.  Higher $M_{HI}/L$'s with lower central surface brightness
reflects the obvious connection between a galaxy's gas supply and the
density of its stellar population.  LSB dwarf galaxies have converted very
little of their gas into stars (i.e. their past star formation has been
inefficient and/or at a very low rate).  Thus, they maintain a high gas
mass value in comparison to an even lower luminosity.

The correlation with galaxy color follows from the surface brightness
correlation.  Though star formation in LSB galaxies occurs at a low rate
(McGaugh 1992), even the smallest amount of star formation has a large
impact on the total galaxy color due to the low surface brightness of the
underlying population.  Galaxies with highest $M_{HI}/L$ values have the
bluest $V-I$ colors, meaning that the portion of the stellar population
that contributes a majority of the optical luminosity in a LSB dwarf
galaxy is quite young (see also Schombert, McGaugh and Eder 2000).

\subsection{Dynamical Masses and Baryonic Matter}

The dynamical mass values are listed in Table 2 for the dwarf galaxies
with optical imaging and the histograms of $M_{dyn}$ and $M_{dyn}/L$
values are found in the of Figure 6.  Note that this ratio is determined
using $I$ band luminosities.  For the typical dwarf color of $B-I=1.5$,
the $M_{dyn}/L$ values are a factor of 4 higher than similar $B$ values
(Staveley-Smith, Davies and Kinman 1992).  In general, the dynamical
masses are a factor of 10 higher than the H\,I masses, but still much
lower than a typical spiral sample (de Jong 1996).  Again, as we have seen
with indirect mass indicators (luminosity and size), galaxies selected by
dwarf-like morphology define a low mass population.  As shown in Figure 6,
LSB dwarfs range in $M_{dyn}/L$ from 1.5 to 50 with a mean of 20.  For
comparison, the mean $M/L$ of spiral galaxy pairs is 40 at radii of 100
kpc (Honma 1999).  

A comparison of dynamical mass to stellar mass (as represented by the
total $I$ band magnitude of the dwarf) is found in Figure 7.  Magnitudes
in the $I$ band are preferred over other bandpasses because stellar
population models (Worthey 1994) show that the stellar mass to luminosity
ratio ($M_*/L$) vary with time and star formation rate as a function of
wavelength, but is most stable in the far-red.  The use of the $I$ band
observations minimizes these stellar populations effects due to its
distance in wavelength from the region around the 4000\AA\ break.  Thus,
$I$ band measurements 1) provide a more accurate estimate of the stellar
mass of a galaxy, 2) obtain a luminosity measure that vary little with
recent star formation and 3) determine structural parameters (such as
scale length) which are undistorted by recent star formation events.

Although there is a great deal of scatter in the $M_T^I$ versus $M_{dyn}$
diagram, the LSB dwarf sample defines a linear correlation with a constant
$M/L$ slope.  Immediately obvious is that none of the dwarfs have their
total masses composed solely of stars.  Nor is the total mass explained by
the sum of the stellar and gas masses (see below).  In other words, LSB
dwarfs are dark matter dominated.  Interestingly, the linear correlation
means that the dark matter follows the stellar light, even for this sample
of extreme $M/L$ objects.  One could interpret this one-to-one
correspondence to indicate that dark matter is baryonic (i.e. baryons
follow baryons).  However, it is equally probable that the dark matter
component is non-baryonic and simply accretes baryon material in a fashion
that is proportional to its own mass.  Certainly, with the discovery of
massive neutrinos (see Turner 1997), it is now known that at least some
fraction of the dark matter component in galaxies is non-baryonic.

The remaining question, regarding the dynamical mass estimates, is how
much of a dwarf galaxies mass is in the form of observable baryonic
material.  This question is answered, graphically, in Figure 8 where we
plot the ratios of stellar mass to dynamical mass and gas mass to
dynamical mass.  The stellar mass is calculated assuming a $M_*/L$ of 1.2,
based on the optical colors of LSB galaxies (McGaugh and de Blok 1997).
Again, the advantage of $I$ band magnitudes is that this value is very
insensitive to recent star formation, varying only 10\% for the range of
observed optical colors in LSB dwarfs.  The gas mass is calculated from
the H\,I mass and multiplying for a factor of 1.4 to account for helium
and metals.  There has been no observed CO emission in LSB galaxies
(Schombert \etal 1990) and the current star formation signature by
H$\alpha$ emission is weak.  Thus, we take the amount of molecular and
ionized gas to be negligible.  Although there has been discussion of
ionization by extragalactic UV (Corbelli and Salpeter 1993), none has been
detected in LSB galaxies.  The sum of the stellar and gas masses produces
the mass in baryons ($M_{baryons}$) for each galaxy shown in Figure 8.

Unlike the rest of the Hubble sequence, LSB dwarfs tend to have equal
amounts of mass in the form of stars and gas.  However, both these
components are a small fraction of the total mass.  The mean for
$M_{stars}/M_{dyn}$ is 0.11 and the mean for $M_{gas}/M_{dyn}$ is 0.08.
This is similar to the value found by Burlak (1996) of
$M_{gas}/M_{stars}=0.7$.  Combining the gas and stellar mass numbers, we
find that the mass in baryons for these dwarf galaxies is less than 20\%
the total mass, on average.  Again, in agreement with a value of 26\% from
Burlak.

\section{CONCLUSIONS}

The paper is the third in a series concerning the properties of newly
discovered LSB dwarfs.  These galaxies were selected by irregular
morphology from deep photographic plates (PSS-II).  While the emphasis was
on their visual appearance, this method to find new, uncataloged galaxies
is automatically biased the search to objects with very low central
surface brightnesses (i.e. not visible on previous sky surveys).  The
combination of low surface brightness and dwarf-like morphology was highly
effective at producing a sample of low mass, low luminosity and small
metric size galaxies.  The results from the H\,I data presented herein can
be summarized as the following:

\begin{itemize}

\item[(1)] Low mass dwarfs exists, but are rare.  Contrary to claims
(Briggs 1997), our current galaxy catalogs are incomplete for objects with
H\,I masses less than $10^7 M_{\sun}$ since these objects are small and
low in surface brightness and would be confused with Galactic emission in
H\,I surveys.

\item[(2)] The H\,I mass distribution has a mean of $M_{HI}=7.9\times10^8
M_{\sun}$ with a sharp cutoff above $M_{HI}=4.0\times10^9 M_{\sun}$ and a
long tail to low H\,I masses.  This distribution is much lower than a
typical spiral sample and demonstrates that, although LSB galaxies are
H\,I-rich, the gas component does not make them high mass.  Mean H\,I
surface densities are below the critical density for star formation based
on the Toomre criteria, justifying statements that LSB dwarfs are
inefficient at star formation and have quiescent past histories in
agreement with their LSB nature.

\item[(3)] LSB dwarfs continue the trend from late-type spirals of high
$M_{HI}/L$, increasing with lower central surface brightnesses and bluer
colors.  The explanation of this correlation lies in a quiescent star
formation history which leaves a large gas reservoir on top of a faint
stellar population.  Any recent star formation, even at a very low rate,
has a sharp blueward effect on the underlying LSB structure.

\item[(4)] The $M_T^I$,$M_{dyn}$ correlation (Figure 7) demonstrates that
the LSB dwarf sample that the dark matter (whether baryonic or
non-baryonic) follows the baryonic matter.  The distribution of
$M_{dyn}/L$ ranges from 1.5 to 50 (note that these are $I$ band numbers,
a factor of 4 higher than equivalent values in the $B$ band). 

\end{itemize}

\acknowledgements
The authors wish to thank the generous support of the Arecibo Observatory
for the allocation of time to search for HI emission from the candidate
dwarf galaxies and Michigan- Dartmouth-M.I.T Observatory in carrying out
the photometry portion of this program This work is based on photographic
plates obtained at the Palomar Observatory 48-inch Oschin Telescope for
the Second Palomar Observatory Sky Survey which was funded by the Eastman
Kodak Company, the National Geographic Society, the Samuel Oschin
Foundation, the Alfred Sloan Foundation, the National Science Foundation
and the National Aeronautics and Space Administration.

\clearpage

\figcaption{H\,I profiles for the LSB dwarf sample.  Baseline fits are
shown as solid lines.}

\figcaption{H\,I flux and profile widths ($W_{50}$) for the LSB dwarf
sample.  Most of the sample has widths below 150 km sec$^{-1}$.  Galaxies
with widths less than 40 km sec$^{-1}$ would be indistinguishable from RFI.}

\figcaption{Size and Velocity distributions for the LSB dwarf sample.  The
top panel displays the histogram scale lengths, $\alpha$, from exponential
fits to the dwarfs surface brightness profiles. For comparison, the mean
scale length of a disk galaxy (de Jong 1996) is indicated.  The bottom
panel displays H\,I mass as a function of redshift.  The solid line is the
3$\sigma$ detection limit for a dwarf galaxy with a line width ($W_{50}$)
of 75 km sec$^{-1}$.  There appears to be no lower limit for H\,I mass for
this sample.}

\figcaption{The distribution of H\,I mass and H\,I surface density for the
LSB dwarf sample.  The sample has a mean value of $M_{HI}=7.9\times10^8
M_{\sun}$ and a sharp cutoff at $M_{HI}=4.0\times10^9 M_{\sun}$.  The
bottom panel displays mean H\,I surface density ($\Sigma_{HI}$) for each
dwarf with available optical imaging.  The critical threshold for star
formation based on the Toomre (1964) instability criterion is between
$9\times10^{20}$ and $6\times10^{21}$ atoms cm$^{-2}$, which corresponds
to the range of 5 and 30 $M_{\sun}$ pc$^{-2}$ (van Zee \etal 1997).}

\figcaption{The H\,I mass to light ratio for the LSB dwarf sample and the
de Jong sample of spirals.  In the top panel, $M_{HI}/L$ is plotted
against galaxy central surface brightness ($\mu_o$).  While the
correlation is poor for the dwarfs, they do continue the spiral trend of
increasing $M_{HI}/L$ with decreasing surface brightness.  The bottom
panel plots $M_{HI}/L$ versus galaxy color ($V-I$).  Galaxies with highest
$M_{HI}/L$ values have the bluest $V-I$ colors, meaning that the portion
of the stellar population that contributes a majority of the optical
luminosity in a LSB dwarf galaxy is quite young.}

\figcaption{The distribution of dynamical mass and total mass to light
ratio for the LSB dwarf sample.  The top panel displays the histogram of
$M_{dyn}$ values for dwarfs with optical imaging to determine scale
length.  The bottom panel displays the histogram for the dynamical (total)
mass to light ratio based on $I$ band magnitudes.}

\figcaption{Dynamical mass versus total magnitude for the subset of the
LSB dwarf sample with optical imaging.  Lines of constant $M/L$ are
marked, as well as the boundary where the mass from stars equals the total
mass (assuming $M_{stars}/L$ = 1.2 (see text).}

\figcaption{Ratios of baryons from stars and gas to total mass for the LSB
dwarf sample.  The top two panels display the histograms of the ratio of
mass from gas and stars (assuming standard corrections for helium, metals
and $M_*/L$, see text).  The bottom panel displays the histogram for the
sum of the baryons from gas and stars divided by total mass.  For a
majority of LSB dwarfs, the mass in visible baryons is less than 20\% the
total mass of the galaxy.}

\pagestyle{empty} 

\baselineskip 14pt \lineskip 2pt \parskip 2pt plus 1 pt
\begin{deluxetable}{lccccccccc}

\tablecolumns{10}
\tablewidth{0pc}
\tablecaption{LSB Dwarf Galaxy Sample}
\tablehead{
\colhead{Field} &
\colhead{RA} &
\colhead{DEC} &
\colhead{V$_{h}$}&
\colhead{V$_{0}$} &
\colhead{W} &
\colhead{D/S} &
\colhead{F.I.} &
\colhead{log $M_{HI}$} &
\colhead{run} \\ 
\colhead{} &
\colhead{1950} &
\colhead{1950} &
\colhead{km s$^{-1}$} &
\colhead{km s$^{-1}$} &
\colhead{km s$^{-1}$} &
\colhead{} &
\colhead{mJy km s$^{-1}$} &
\colhead{$M_{\sun}$} &
\colhead{} \\
\nl
\colhead{(1)} &
\colhead{(2)} &
\colhead{(3)} &
\colhead{(4)} &
\colhead{(5)} &
\colhead{(6)} &
\colhead{(7)} &
\colhead{(8)} &
\colhead{(9)} &
\colhead{(10)} 
} 
\startdata
409-1&000022.2&291821&4836&5074&114&D&\phm{1}1400&9.21&1\nl
409-3&000204.9&291914&\nodata&\nodata&\nodata&\nodata&\nodata&\nodata&3\nl
409-4&000328.4&270417&3181&3412&\phm{1}69&D:&\phm{1}1500&8.90&1\nl
409-5&000452.7&314739&4864&5105&100&D&\phm{1}2100&8.90&1\nl
409-6&000729.4&275704&\nodata&\nodata&\nodata&\nodata&\nodata&\nodata&3\nl
409-7&001141.3&283710&7060&7291&167&:&\phm{1}1800&9.63&1\nl
409-9&001256.3&282244&6682&6911&\phm{1}87&D&\phm{13}930&9.30&1\nl
409-11&001349.5&271754&4597&4823&\phm{1}92&D&\phm{13}870&8.96&1\nl
409-13&234841.0&324243&4951&5202&128&D&\phm{13}940&9.06&1\nl
409-14&235711.5&265838&4699&4934&\phm{1}64&S&\phm{13}890&8.99&1\nl
409-15&000639.0&301218&\nodata&\nodata&\nodata&\nodata&\nodata&\nodata&3\nl
409-16&235004.7&305748&5161&5410&\phm{1}68&S&\phm{13}900&9.07&3\nl
409-17&235557.0&302336&4811&5060&\phm{1}68&S&\phm{13}523&8.78&3\nl
409-18&001348.5&301200&6904&7135&155&D&\phm{1}1835&9.62&3\nl
409-19&001033.5&293730&7039&7270&138&:&\phm{13}638&9.18&3\nl
409-20&001053.3&293648&\nodata&\nodata&\nodata&\nodata&\nodata&\nodata&3\nl
\nl
474-1&003221.5&243454&5480&5689&121&D&\phm{13}610&8.95&1\nl
474-4&004118.1&255225&4898&5105&\phm{1}38&S&\phm{13}950&9.05&1\nl
474-5&004322.9&273108&\nodata&\nodata&\nodata&\nodata&\nodata&\nodata&3\nl
474-7&004731.3&265708&4922&5123&119&D&\phm{13}905&9.03&3\nl
474-9&005254.7&235529&4980&5174&\phm{1}82&D&\phm{1}1300&9.19&1\nl
\nl
475-3&010146.0&221330&5662&5845&118&D&\phm{1}1200&9.27&1\nl
475-4&011201.2&265223&3618&3807&111&D&\phm{1}4900&9.50&1\nl
\nl
476-2&012118.1&264715&4908&5091&163&D&\phm{1}2800&9.51&1\nl
476-3&012322.4&274231&4021&4205&\phm{1}75&D:&\phm{1}1100&8.94&1\nl
476-4&012722.2&253628&3678&3856&152&D&\phm{1}9300&9.79&1\nl
476-5&012837.2&234149&3414&3582&\phm{1}65&S&\phm{1}4400&9.40&1\nl
476-6&013428.5&261057&3872&4044&\phm{1}66&D:&\phm{13}990&8.86&1\nl
476-8&013929.3&260657&\phm{1}359&\phm{1}527&\phm{1}38&S&\phm{1}1100&7.14&1\nl
476-9&014009.0&271848&\nodata&\nodata&\nodata&\nodata&\nodata&\nodata&3\nl
\nl
495-1&081921.5&224606&2309&2215&\phm{1}75&D&\phm{13}389&7.93&2\nl
495-2&082657.7&253930&2197&2116&\phm{1}34&S&\phm{13}669&8.13&2\nl
495-3&082518.0&255724&2267&2188&\phm{1}50&S&\phm{13}546&8.07&2\nl
\nl
500-1&102116.1&263812&\nodata&\nodata&\nodata&\nodata&\nodata&\nodata&2\nl
500-3&100311.3&240642&1347&1260&\phm{1}92&D&\phm{1}1048&7.87&2\nl
500-4&101633.5&225712&1576&1486&\phm{1}67&S&\phm{13}395&7.59&2\nl
500-5&101551.0&215724&4078&3983&178&D&\phm{1}2802&9.30&2\nl
500-6&102513.7&222624&9346&9256&\phm{1}70&S&\phm{13}695&9.43&2\nl
\nl
508-1&130325.8&275842&2527&2539&228&D&\phm{1}6128&9.25&2\nl
508-2&130208.4&270248&1924&1931&\phm{1}84&S&\phm{1}3104&8.72&2\nl
508-3&130032.4&262054&\nodata&\nodata&\nodata&\nodata&\nodata&\nodata&2\nl
508-4&132152.8&253742&\nodata&\nodata&\nodata&\nodata&\nodata&\nodata&2\nl
508-5&131049.8&243706&4406&4408&\phm{1}85&S&\phm{13}835&8.86&2\nl
\nl
512-1&143436.0&275124&7199&7276&156&D&\phm{1}1082&9.41&2\nl
512-2&143108.4&271300&\phm{1}831&\phm{1}903&\phm{1}75&S&\phm{1}2391&7.94&2\nl
512-3&142851.0&264018&4201&4269&119&D:&\phm{13}761&8.80&2\nl
512-4&143003.0&263236&4805&4874&102&D&\phm{1}2544&9.43&2\nl
512-5&142954.6&260424&\nodata&\nodata&\nodata&\nodata&\nodata&\nodata&2\nl
512-6&144321.6&255348&4269&4345&119&D&\phm{1}3717&9.50&2\nl
512-7&144423.4&255448&4261&4338&127&D&\phm{1}2452&9.32&2\nl
512-8&142605.4&251612&\nodata&\nodata&\nodata&\nodata&\nodata&\nodata&2\nl
512-9&143744.4&243424&4499&4566&119&D&\phm{1}1819&9.23&2\nl
512-10&143630.6&232906&4526&4588&\phm{1}76&S&\phm{13}528&8.70&2\nl
512-11&144631.2&231200&\nodata&\nodata&\nodata&\nodata&\nodata&\nodata&2\nl
512-12&144404.8&225312&\nodata&\nodata&\nodata&\nodata&\nodata&\nodata&2\nl
\nl
514-1&151929.4&275706&\nodata&\nodata&\nodata&\nodata&\nodata&\nodata&2\nl
514-2&152820.4&272612&2023&2139&\phm{1}33&S&\phm{1}1591&8.52&2\nl
514-3&151745.6&262842&\nodata&\nodata&\nodata&\nodata&\nodata&\nodata&2\nl
514-4&151753.4&262848&\nodata&\nodata&\nodata&\nodata&\nodata&\nodata&2\nl
514-5&152146.2&242336&6579&6680&\phm{1}69&D&\phm{13}573&9.06&2\nl
514-6&152618.6&243406&\nodata&\nodata&\nodata&\nodata&\nodata&\nodata&2\nl
\nl
516-1&161228.8&272206&\nodata&\nodata&\nodata&\nodata&\nodata&\nodata&2\nl
516-2&160836.0&255012&9767&9910&132&D&\phm{13}610&9.43&2\nl
516-3&160557.0&242612&9750&9886&\phm{1}35&S&\phm{13}672&9.47&2\nl
516-4&160930.6&241542&9322&9460&\phm{1}44&S&\phm{13}370&9.17&2\nl
\nl
535-1&225457.9&274242&2944&3207&108&D&\phm{1}2400&9.05&1\nl
535-2&225636.1&260907&7705&7964&\phm{1}65&D&\phm{13}840&9.38&1\nl
535-3&225702.6&245132&7417&7673&\phm{1}75&D:&\phm{13}730&9.29&1\nl
535-4&230114.7&232459&1155&1406&\phm{1}94&D&\phm{1}2700&8.38&1\nl
535-5&230354.6&273713&6913&7173&\phm{1}68&S&\phm{13}530&9.09&1\nl
535-6&230552.2&272647&2884&3143&\phm{1}86&D&\phm{1}2900&9.11&1\nl
535-8&231651.8&243937&5809&6057&\phm{1}94&D&\phm{1}1500&9.39&1\nl
535-9&231913.4&272316&5935&6188&\phm{1}45&S:&\phm{13}940&9.21&1\nl
\nl
538-1&000014.4&184122&\phm{1}713&\phm{1}924&\phm{1}50&S&\phm{1}1100&7.63&1\nl
538-2&000156.8&202620&2175&2383&\phm{1}84&S:&\phm{1}1977&8.70&3\nl
538-3&000213.0&180156&6315&6523&\phm{1}77&D:&\phm{1}1000&9.28&1\nl
538-4&000325.9&213506&\nodata&\nodata&\nodata&\nodata&\nodata&\nodata&3\nl
538-7&000839.5&225554&\nodata&\nodata&\nodata&\nodata&\nodata&\nodata&3\nl
538-11&001021.1&181543&5514&5718&\phm{1}32&S&\phm{13}490&8.86&1\nl
538-12&001118.3&230333&\nodata&\nodata&\nodata&\nodata&\nodata&\nodata&3\nl
538-15&235256.9&173829&1769&1981&150&D&15000&9.42&1\nl
538-16&235312.9&180853&1739&1953&\phm{1}94&D&\phm{1}6200&9.03&1\nl
538-17&235706.6&172137&6457&6666&174&D&\phm{1}1200&9.38&1\nl
538-18&235903.9&171238&6413&6621&\phm{1}72&S:&\phm{13}850&9.22&1\nl
\nl
539-1&001428.9&224311&4405&4619&118&D&\phm{1}2200&9.32&1\nl
539-2&001440.6&171450&1005&1204&\phm{1}72&D&\phm{1}2200&8.16&1\nl
539-3&001444.1&203136&4366&4574&144&D&\phm{1}5000&9.67&1\nl
539-4&001613.7&194245&5542&5747&\phm{1}72&S&\phm{1}1400&9.32&1\nl
539-5&001805.3&221855&7352&7563&197&D&\phm{1}3500&9.95&1\nl
539-6&001822.7&172447&5235&5432&136&D&\phm{1}1500&9.30&1\nl
539-8&002152.2&204642&5246&5451&\phm{1}76&D&\phm{13}580&8.89&1\nl
\nl
540-1&002958.9&221817&4560&4764&118&D&\phm{1}1100&9.05&1\nl
540-2&003804.7&222406&5899&6098&\phm{1}73&D&\phm{13}660&9.04&1\nl
540-3&004333.6&191221&2646&2832&231&D&18000&9.81&1\nl
540-4&004417.8&211818&5180&5372&\phm{1}72&S&\phm{1}1200&9.19&1\nl
540-8&005404.3&225008&2594&2778&\phm{1}92&D&\phm{1}1169&8.61&3\nl
\nl
541-4&010559.6&211529&5546&5723&100&D&\phm{1}1700&9.40&1\nl
\nl
559-1&071941.3&172112&\nodata&\nodata&\nodata&\nodata&\nodata&\nodata&2\nl
559-2&072650.3&191330&\nodata&\nodata&\nodata&\nodata&\nodata&\nodata&2\nl
559-3&073215.0&191830&8580&8482&576&D&\phm{1}4144&10.13&2\nl
559-4&071904.7&205054&\nodata&\nodata&\nodata&\nodata&\nodata&\nodata&2\nl
559-5&071515.5&211054&\nodata&\nodata&\nodata&\nodata&\nodata&\nodata&2\nl
559-6&072218.5&221048&\nodata&\nodata&\nodata&\nodata&\nodata&\nodata&2\nl
\nl
561-1&075748.0&200006&7714&7612&\phm{1}95&D&\phm{1}1585&9.62&2\nl
561-2&080345.0&190736&4631&4523&\phm{1}85&D&\phm{13}866&8.90&2\nl
\nl
563-1&085319.1&172706&4161&4037&\phm{1}34&S&\phm{13}531&8.59&2\nl
563-2&085012.5&183742&4319&4201&119&D&\phm{1}3928&9.49&2\nl
563-3&084348.0&190430&3935&3820&\phm{1}85&D:&\phm{13}940&8.79&2\nl
563-4&085215.0&195630&3495&3384&110&D&\phm{1}5097&9.42&2\nl
563-5&083357.5&195042&4645&4535&110&D&\phm{1}2700&9.40&2\nl
563-6&084936.5&211230&3567&3462&110&D&\phm{13}321&8.24&2\nl
\nl
564-1&085455.1&224942&\nodata&\nodata&\nodata&\nodata&\nodata&\nodata&2\nl
564-2&091146.1&230000&8131&8034&\phm{1}70&D&\phm{13}231&8.83&2\nl
564-4&090603.5&213400&2703&2599&109&D&\phm{13}557&8.23&2\nl
564-5&085417.3&210048&\nodata&\nodata&\nodata&\nodata&\nodata&\nodata&2\nl
564-6&085628.7&204536&\nodata&\nodata&\nodata&\nodata&\nodata&\nodata&2\nl
564-7&090227.5&204148&\nodata&\nodata&\nodata&\nodata&\nodata&\nodata&2\nl
564-8&090001.7&201630&\phm{1}488&\phm{1}378&\phm{1}58&S&\phm{1}1819&7.07&2\nl
564-10&090628.1&183536&\nodata&\nodata&\nodata&\nodata&\nodata&\nodata&2\nl
564-11&090423.3&180300&4284&4163&153&D&\phm{13}952&8.87&2\nl
564-12&085723.3&174648&3840&3718&144&D&\phm{1}2266&9.15&2\nl
564-13&090027.5&173318&4292&4169&153&D:&\phm{13}433&8.53&2\nl
564-14&090547.3&174218&\nodata&\nodata&\nodata&\nodata&\nodata&\nodata&2\nl
564-15&090830.5&172242&3386&3262&135&D&\phm{1}1125&8.73&2\nl
564-16&090825.1&201054&6873&6762&\phm{1}35&D&\phm{13}237&8.69&2\nl
564-17&085408.3&203106&\nodata&\nodata&\nodata&\nodata&\nodata&\nodata&2\nl
\nl
565-1&091509.0&224912&6239&6141&103&D&\phm{1}1092&9.27&2\nl
565-2&092139.0&222918&3830&3731&110&D&\phm{1}6472&9.61&2\nl
565-3&093448.5&221524&7623&7523&165&D&\phm{13}999&9.41&2\nl
565-4&093326.3&215212&\nodata&\nodata&\nodata&\nodata&\nodata&\nodata&2\nl
565-5&092958.7&214118&\phm{1}558&\phm{1}455&364&D&\phm{1}4659&7.64&2\nl
565-7&093657.5&210906&\nodata&\nodata&\nodata&\nodata&\nodata&\nodata&2\nl
565-8&093822.7&205500&\nodata&\nodata&\nodata&\nodata&\nodata&\nodata&2\nl
565-9&093024.5&204430&\nodata&\nodata&\nodata&\nodata&\nodata&\nodata&2\nl
565-10&092724.0&201300&\phm{1}561&\phm{1}451&\phm{1}33&S&\phm{13}732&6.83&2\nl
565-11&093117.3&200518&\nodata&\nodata&\nodata&\nodata&\nodata&\nodata&2\nl
565-12&091516.1&194236&\nodata&\nodata&\nodata&\nodata&\nodata&\nodata&2\nl
565-13&091842.0&193636&\nodata&\nodata&\nodata&\nodata&\nodata&\nodata&2\nl
565-14&091725.1&180648&\nodata&\nodata&\nodata&\nodata&\nodata&\nodata&2\nl
565-15&093134.7&180448&\nodata&\nodata&\nodata&\nodata&\nodata&\nodata&2\nl
565-16&093710.7&173412&\nodata&\nodata&\nodata&\nodata&\nodata&\nodata&2\nl
\nl
568-2&104214.4&210730&1219&1128&\phm{1}33&S&\phm{13}594&7.53&2\nl
568-3&102118.0&212030&5731&5635&128&D:&\phm{1}1604&9.36&2\nl
568-4&101812.0&202424&1557&1455&\phm{1}75&D&\phm{1}1293&8.09&2\nl
568-5&103006.0&193324&\nodata&\nodata&\nodata&\nodata&\nodata&\nodata&2\nl
\nl
570-1&112234.1&170936&1021&\phm{1}927&\phm{1}50&S&\phm{13}783&7.48&2\nl
570-2&112225.7&172148&\nodata&\nodata&\nodata&\nodata&\nodata&\nodata&2\nl
570-3&112022.1&174448&1383&1290&\phm{1}67&D&\phm{13}465&7.54&2\nl
570-4&111542.0&175812&1057&\phm{1}963&\phm{1}66&S&\phm{1}1860&7.89&2\nl
570-5&111333.5&181336&\nodata&\nodata&\nodata&\nodata&\nodata&\nodata&2\nl
570-6&112103.5&193254&1894&1810&\phm{1}92&D&\phm{13}189&7.44&2\nl
570-7&111938.3&210900&\phm{1}810&\phm{1}733&\phm{1}50&S&\phm{1}1066&7.41&2\nl
570-8&110057.5&224742&\nodata&\nodata&\nodata&\nodata&\nodata&\nodata&2\nl
\nl
571-1&113436.0&212948&\nodata&\nodata&\nodata&\nodata&\nodata&\nodata&2\nl
571-2&114551.6&203954&6796&6729&\phm{1}60&S&\phm{13}888&9.26&2\nl
571-3&113548.6&201530&6981&6907&147&D:&\phm{1}1331&9.46&2\nl
571-4&112022.2&194524&\nodata&\nodata&\nodata&\nodata&\nodata&\nodata&2\nl
571-5&112104.8&193300&6132&6048&\phm{1}77&D:&\phm{13}230&8.58&2\nl
571-6&112613.8&183324&\nodata&\nodata&\nodata&\nodata&\nodata&\nodata&2\nl
571-7&113452.2&183542&\nodata&\nodata&\nodata&\nodata&\nodata&\nodata&2\nl
571-8&113134.8&172542&\nodata&\nodata&\nodata&\nodata&\nodata&\nodata&2\nl
\nl
572-1&114815.5&172518&\nodata&\nodata&\nodata&\nodata&\nodata&\nodata&2\nl
572-2&115201.7&175642&3746&3670&144&D:&\phm{13}721&8.64&2\nl
572-3&114855.7&173812&\nodata&\nodata&\nodata&\nodata&\nodata&\nodata&2\nl
572-4&120458.1&185754&\phm{1}793&\phm{1}729&\phm{1}50&S&\phm{13}877&7.32&2\nl
572-5&114539.2&185536&\phm{1}976&\phm{1}901&\phm{1}66&S&\phm{1}2198&7.90&2\nl
572-6&115532.3&210700&4074&4014&\phm{1}59&S&\phm{1}1743&9.10&2\nl
572-7&115225.1&210012&\nodata&\nodata&\nodata&\nodata&\nodata&\nodata&2\nl
\nl
575-1&124918.6&220042&\phm{1}580&\phm{1}557&\phm{1}33&S&\phm{1}2386&7.52&2\nl
575-2&124953.4&215430&\phm{1}764&\phm{1}741&116&D:&\phm{1}8329&8.31&2\nl
575-3&130016.2&221554&\nodata&\nodata&\nodata&\nodata&\nodata&\nodata&2\nl
575-4&130010.1&221454&\nodata&\nodata&\nodata&\nodata&\nodata&\nodata&2\nl
575-5&125313.2&192848&\phm{1}419&\phm{1}387&\phm{1}48&S&\phm{1}4373&7.47&2\nl
575-6&124549.2&185442&\nodata&\nodata&\nodata&\nodata&\nodata&\nodata&2\nl
575-7&125607.2&180454&1013&\phm{1}977&\phm{1}75&S&\phm{1}2463&8.02&2\nl
575-8&130201.2&180742&\nodata&\nodata&\nodata&\nodata&\nodata&\nodata&2\nl
575-9&130414.4&181600&\nodata&\nodata&\nodata&\nodata&\nodata&\nodata&2\nl
\nl
576-1&132110.7&221006&7096&7095&216&D&\phm{1}1541&9.54&2\nl
576-2&132549.8&210736&\nodata&\nodata&\nodata&\nodata&\nodata&\nodata&2\nl
576-3&131146.1&204700&6504&6491&\phm{1}78&S&\phm{13}650&9.09&2\nl
576-4&131419.7&203254&\nodata&\nodata&\nodata&\nodata&\nodata&\nodata&2\nl
576-5&131407.7&203142&\nodata&\nodata&\nodata&\nodata&\nodata&\nodata&2\nl
576-6&130733.5&195906&3731&3711&110&D&\phm{1}3941&9.39&2\nl
576-7&131344.3&190136&6553&6534&\phm{1}69&D:&\phm{13}720&9.14&2\nl
576-8&132219.7&192530&7165&7154&190&D&\phm{1}2202&9.70&2\nl
576-9&131255.7&181554&3831&3808&110&D&\phm{1}1262&8.92&2\nl
576-10&132914.3&171724&\nodata&\nodata&\nodata&\nodata&\nodata&\nodata&2\nl
576-11&132424.0&201400&7168&7161&\phm{1}52&D&\phm{1}2559&9.77&2\nl
\nl
577-1&134510.2&204142&\nodata&\nodata&\nodata&\nodata&\nodata&\nodata&2\nl
577-2&133457.0&193106&6545&6543&\phm{1}95&D&\phm{1}1410&9.43&2\nl
577-3&133040.2&183936&6936&6928&121&D&\phm{13}643&9.14&2\nl
577-4&133355.8&182448&7080&7073&\phm{1}78&S&\phm{1}1388&9.49&2\nl
577-5&133815.0&172054&4095&4087&110&D&\phm{1}1541&9.06&2\nl
577-6&134625.2&173200&7572&7571&\phm{1}61&S&\phm{13}602&9.19&2\nl
\nl
582-1&153153.4&215754&7065&7166&276&D&\phm{1}2330&9.73&2\nl
582-2&152340.8&204824&4575&4665&\phm{1}51&D&\phm{13}273&8.43&2\nl
582-3&153319.8&204824&\nodata&\nodata&\nodata&\nodata&\nodata&\nodata&2\nl
582-4&153528.2&205700&4465&4565&\phm{1}85&D&\phm{13}302&8.45&2\nl
582-5&151559.4&184918&\nodata&\nodata&\nodata&\nodata&\nodata&\nodata&2\nl
\nl
584-1&160931.1&170030&4669&4783&\phm{1}85&D&\phm{1}1533&9.20&2\nl
584-2&155935.3&185200&2628&2741&109&D&\phm{1}2051&8.84&2\nl
584-3&161520.3&193854&2576&2704&126&D&\phm{1}1746&8.76&2\nl
584-4&155515.5&204842&2269&2385&167&D&\phm{1}7885&9.30&2\nl
584-5&161913.1&205900&3104&3239&135&D&\phm{1}8027&9.58&2\nl
584-6&160613.7&211918&4787&4913&\phm{1}85&S&\phm{1}2147&9.37&2\nl
584-7&155816.7&211824&\nodata&\nodata&\nodata&\nodata&\nodata&\nodata&2\nl
\nl
609-2&003545.3&131220&5597&5770&102&:&\phm{13}660&8.99&1\nl
609-3&003548.3&131244&5459&5632&158&:&\phm{1}1600&9.36&1\nl
609-4&004217.2&163957&4678&4845&196&D&\phm{1}2234&9.37&3\nl
609-5&005210.2&132325&5428&5590&\phm{1}66&S&\phm{1}1300&9.26&1\nl
\nl
611-5&011706.5&163148&2163&2318&\phm{1}32&S&\phm{1}1200&8.46&1\nl
611-6&011823.7&164844&2495&2650&142&D&\phm{1}4381&9.14&3\nl
611-7&011838.9&120902&\phm{1}644&\phm{1}783&103&D:&12000&8.52&1\nl
611-10&012038.9&170143&5806&5961&214&D&\phm{1}1326&9.33&3\nl
611-11&012235.0&135352&5041&5183&\phm{1}35&S&\phm{1}1200&9.16&1\nl
611-13&012432.9&132036&4505&4643&102&D&\phm{1}2500&9.38&1\nl
611-14&012606.4&162555&\phm{1}594&\phm{1}742&\phm{1}49&S&\phm{1}2200&7.74&1\nl
611-15&012725.5&132031&6395&6531&120&D&\phm{1}1800&9.54&1\nl
611-16&012734.8&142511&2449&2589&121&D&\phm{1}3800&9.06&1\nl
611-17&012855.9&130429&2739&2873&107&D&\phm{13}820&8.48&1\nl
611-18&012905.9&135354&6763&6900&175&D&\phm{1}2300&9.69&1\nl
611-21&013125.1&123501&5019&5149&104&D&\phm{13}860&9.01&1\nl
\nl
631-1&075950.3&160900&4698&4577&136&D&\phm{1}1593&9.18&2\nl
631-2&080813.1&151400&\nodata&\nodata&\nodata&\nodata&\nodata&\nodata&2\nl
631-3&080246.7&145448&\nodata&\nodata&\nodata&\nodata&\nodata&\nodata&2\nl
631-4&075945.5&144812&\nodata&\nodata&\nodata&\nodata&\nodata&\nodata&2\nl
631-5&075731.1&145054&\nodata&\nodata&\nodata&\nodata&\nodata&\nodata&2\nl
631-6&075405.4&144630&\nodata&\nodata&\nodata&\nodata&\nodata&\nodata&2\nl
631-7&075413.7&143118&\phm{1}343&\phm{1}216&\phm{1}83&S&15390&7.51&2\nl
631-8&075707.7&135848&4471&4340&127&D&\phm{13}929&8.90&2\nl
631-9&075439.0&124018&\nodata&\nodata&\nodata&\nodata&\nodata&\nodata&2\nl
631-10&075653.3&121248&\nodata&\nodata&\nodata&\nodata&\nodata&\nodata&2\nl
631-11&080854.0&120536&\nodata&\nodata&\nodata&\nodata&\nodata&\nodata&2\nl
\nl
634-1&085429.3&131448&\nodata&\nodata&\nodata&\nodata&\nodata&\nodata&2\nl
634-2&090130.0&134024&\nodata&\nodata&\nodata&\nodata&\nodata&\nodata&2\nl
634-3&090607.7&144706&\phm{1}318&\phm{1}181&\phm{1}33&S&\phm{13}217&5.50&2\nl
634-5&090039.0&155048&\nodata&\nodata&\nodata&\nodata&\nodata&\nodata&2\nl
634-6&085718.0&162348&\nodata&\nodata&\nodata&\nodata&\nodata&\nodata&2\nl
634-7&085018.5&164848&\nodata&\nodata&\nodata&\nodata&\nodata&\nodata&2\nl
\nl
637-1&095013.1&172348&\nodata&\nodata&\nodata&\nodata&\nodata&\nodata&2\nl
637-2&095431.1&175948&\nodata&\nodata&\nodata&\nodata&\nodata&\nodata&2\nl
637-3&095857.5&175554&2227&2110&\phm{1}92&D&\phm{1}1538&8.49&2\nl
637-4&095054.5&163400&\nodata&\nodata&\nodata&\nodata&\nodata&\nodata&2\nl
637-5&095458.7&160742&7727&7600&148&D&\phm{13}977&9.40&2\nl
637-6&095806.0&154506&\nodata&\nodata&\nodata&\nodata&\nodata&\nodata&2\nl
637-7&095328.7&155054&4580&4452&136&D&\phm{1}2330&9.32&2\nl
637-8&100049.7&161600&7818&7693&\phm{1}87&D&\phm{1}1063&9.45&2\nl
637-9&095913.8&160012&\nodata&\nodata&\nodata&\nodata&\nodata&\nodata&2\nl
637-10&100914.3&163106&\nodata&\nodata&\nodata&\nodata&\nodata&\nodata&2\nl
637-11&100954.0&154648&\nodata&\nodata&\nodata&\nodata&\nodata&\nodata&2\nl
637-12&100642.0&151800&\nodata&\nodata&\nodata&\nodata&\nodata&\nodata&2\nl
637-13&095858.2&151124&\nodata&\nodata&\nodata&\nodata&\nodata&\nodata&2\nl
637-14&100031.1&145224&\nodata&\nodata&\nodata&\nodata&\nodata&\nodata&2\nl
637-15&100701.7&142018&10772\phm{1}&10639\phm{1}&142&D&\phm{1}1313&9.83&2\nl
637-16&100949.1&135948&11401\phm{1}&11267\phm{1}&\phm{1}71&D:&\phm{13}967&9.74&2\nl
637-17&085524.5&140118&7492&7352&113&D&\phm{1}1850&9.65&2\nl
637-18&100454.0&132106&2740&2602&\phm{1}42&S&\phm{1}1469&8.65&2\nl
637-19&095250.3&130136&\nodata&\nodata&\nodata&\nodata&\nodata&\nodata&2\nl
637-20&095927.5&123636&6497&6355&\phm{1}86&D&\phm{1}2067&9.57&2\nl
637-22&095209.0&120036&\nodata&\nodata&\nodata&\nodata&\nodata&\nodata&2\nl
\nl
640-2&105049.2&170306&1093&\phm{1}985&\phm{1}91&S&18530&8.91&2\nl
640-3&105049.2&170306&\nodata&\nodata&\nodata&\nodata&\nodata&\nodata&2\nl
640-4&105049.2&170306&\nodata&\nodata&\nodata&\nodata&\nodata&\nodata&2\nl
640-5&110937.8&160148&\nodata&\nodata&\nodata&\nodata&\nodata&\nodata&2\nl
640-6&104737.2&161424&\nodata&\nodata&\nodata&\nodata&\nodata&\nodata&2\nl
640-7&104758.8&160148&1115&1001&100&D&\phm{1}3764&8.23&2\nl
640-8&105813.8&140918&\nodata&\nodata&\nodata&\nodata&\nodata&\nodata&2\nl
640-9&110333.6&130636&\nodata&\nodata&\nodata&\nodata&\nodata&\nodata&2\nl
640-10&104742.0&133200&\nodata&\nodata&\nodata&\nodata&\nodata&\nodata&2\nl
640-11&104822.8&133542&6783&6658&155&D:&\phm{13}231&8.66&2\nl
640-12&105317.4&123624&\nodata&\nodata&\nodata&\nodata&\nodata&\nodata&2\nl
640-13&105337.2&121642&\phm{1}990&\phm{1}861&\phm{1}42&S&\phm{1}1218&7.61&2\nl
640-14&105533.6&121548&\nodata&\nodata&\nodata&\nodata&\nodata&\nodata&2\nl
640-15&110808.4&121718&3244&3121&211&D&\phm{1}4534&9.30&2\nl
640-16&105225.2&142154&\nodata&\nodata&\nodata&\nodata&\nodata&\nodata&2\nl
\nl
646-1&130206.0&180148&\nodata&\nodata&\nodata&\nodata&\nodata&\nodata&2\nl
646-2&130727.6&172536&3580&3549&135&D&3311&9.27&2\nl
646-3&125301.8&160018&\nodata&\nodata&\nodata&\nodata&\nodata&\nodata&2\nl
646-4&125355.8&152130&\nodata&\nodata&\nodata&\nodata&\nodata&\nodata&2\nl
646-5&125030.0&144024&1044&990&91&D&2496&8.04&2\nl
646-6&125026.4&143942&\nodata&\nodata&\nodata&\nodata&\nodata&\nodata&2\nl
646-7&125609.6&142912&213&162&33&S&6968&6.92&2\nl
646-8&124933.0&131042&2101&2040&50&S:&117&7.34&2\nl
646-9&125042.0&125424&1817&1755&50&S&1452&8.30&2\nl
646-10&125924.0&131136&\nodata&\nodata&\nodata&\nodata&\nodata&\nodata&2\nl
646-11&125412.0&121230&565&503&58&S&2025&7.36&2\nl
\nl
648-1&134239.5&124048&\nodata&\nodata&\nodata&\nodata&\nodata&\nodata&2\nl
648-2&134451.0&133300&\nodata&\nodata&\nodata&\nodata&\nodata&\nodata&2\nl
648-3&135222.1&171954&6990&6993&181&D&1601&9.55&2\nl
648-4&134624.0&173200&7575&7574&69&S&482&9.09&2\nl
648-6&133427.5&173024&\nodata&\nodata&\nodata&\nodata&\nodata&\nodata&2\nl
\nl
651-1&142737.2&175436&\nodata&\nodata&\nodata&\nodata&\nodata&\nodata&2\nl
651-2&142909.6&172524&\nodata&\nodata&\nodata&\nodata&\nodata&\nodata&2\nl
651-3&144612.6&163842&\nodata&\nodata&\nodata&\nodata&\nodata&\nodata&2\nl
651-4&145123.4&160212&6299&6345&129&D&1162&9.32&2\nl
651-5&144740.8&152936&\nodata&\nodata&\nodata&\nodata&\nodata&\nodata&2\nl
651-6&144306.0&135648&\nodata&\nodata&\nodata&\nodata&\nodata&\nodata&2\nl
651-7&144203.0&124248&\nodata&\nodata&\nodata&\nodata&\nodata&\nodata&2\nl
651-8&144404.2&121924&\nodata&\nodata&\nodata&\nodata&\nodata&\nodata&2\nl
\nl
656-1&162554.0&174500&4530&4660&60&S&1333&9.11&2\nl
656-2&161104.2&171918&1090&1207&100&D&4107&8.43&2\nl
656-4&161405.4&161242&\nodata&\nodata&\nodata&\nodata&\nodata&\nodata&2\nl
656-5&161527.6&154906&8950&9065&201&D&970&9.55&2\nl
656-6&161419.2&142942&\nodata&\nodata&\nodata&\nodata&\nodata&\nodata&2\nl
\nl
681-1&002859.6&085342&2156&2319&108&D&5900&9.15&1\nl
681-4&004002.0&094350&5071&5229&63&D:&460&8.75&1\nl
681-5&004035.6&083256&5363&5517&110&D:&660&8.96&1\nl
681-8&004822.4&072543&5274&5428&128&D&492&8.81&3\nl
\nl
685-1&014801.1&115434&6873&6988&166&D&1900&9.62&1\nl
685-4&014906.0&113919&4839&4952&188&D&4600&9.71&1\nl
685-6&015049.2&123451&4656&4771&84&D&630&8.81&1\nl
685-7&015206.7&101908&6085&6191&37&S&2200&9.58&1\nl
685-8&015220.7&103641&6205&6312&193&D&1600&9.46&1\nl
685-11&015427.1&111712&3470&3578&112&D:&2800&9.21&1\nl
685-14&015702.8&101032&6302&6403&77&D:&1100&9.31&1\nl
685-15&015703.4&120251&7828&7936&145&D&1500&9.63&1\nl
685-16&015822.0&081524&7804&7897&147&D&1900&9.73&1\nl
685-18&015919.6&112922&4643&4747&136&D&1200&9.09&1\nl
685-19&020009.4&073432&5727&5816&94&D&806&9.09&1\nl
685-20&020550.3&120608&\nodata&\nodata&\nodata&\nodata&\nodata&\nodata&3\nl
685-21&020940.8&125817&2224&2325&50&S:&917&8.35&1\nl
685-22&021222.4&083553&1590&1673&38&S&1000&8.10&1\nl
\nl
702-1&072816.7&080624&1879&1732&84&D:&2774&8.57&2\nl
702-2&073210.7&110154&\nodata&\nodata&\nodata&\nodata&\nodata&\nodata&2\nl
\nl
704-1&082334.1&111200&\nodata&\nodata&\nodata&\nodata&\nodata&\nodata&2\nl
704-2&082048.5&101236&4236&4083&170&D&2756&9.32&2\nl
704-3&081924.5&082730&4174&4014&76&S:&1169&8.93&2\nl
704-4&081633.5&120236&\nodata&\nodata&\nodata&\nodata&\nodata&\nodata&2\nl
\nl
709-2&100139.5&123730&\nodata&\nodata&\nodata&\nodata&\nodata&\nodata&2\nl
709-3&095902.3&114742&7169&7023&138&D&538&9.08&2\nl
709-4&095827.0&111612&\nodata&\nodata&\nodata&\nodata&\nodata&\nodata&2\nl
709-5&095007.7&112336&9472&9323&88&D:&530&9.32&2\nl
709-6&095551.5&102518&5228&5075&77&D&300&8.54&2\nl
709-7&100454.5&101300&7561&7398&69&S&678&9.22&2\nl
709-8&095846.7&091724&\nodata&\nodata&\nodata&\nodata&\nodata&\nodata&2\nl
709-9&095018.5&081548&2642&2479&126&D&3350&8.97&2\nl
709-10&095023.3&080106&2594&2430&34&S&420&8.05&2\nl
709-11&095657.0&120048&2993&2848&59&S:&221&7.91&2\nl
709-12&095207.1&120012&\nodata&\nodata&\nodata&\nodata&\nodata&\nodata&2\nl
\nl
721-1a&135754.6&123748&5722&5710&111&D&888&9.11&2\nl
721-1&135754.6&123748&6109&6097&69&D&1582&9.42&2\nl
721-2&135940.2&112148&\nodata&\nodata&\nodata&\nodata&\nodata&\nodata&2\nl
721-3&141106.6&110718&\nodata&\nodata&\nodata&\nodata&\nodata&\nodata&2\nl
721-4&140028.8&105148&\nodata&\nodata&\nodata&\nodata&\nodata&\nodata&2\nl
721-5&140017.4&101400&5815&5795&94&D&964&9.16&2\nl
721-7&140359.4&093600&7205&7186&69&S:&1637&9.58&2\nl
721-8&135836.6&092054&5433&5408&103&D:&1036&9.13&2\nl
721-9&135833.0&090100&\nodata&\nodata&\nodata&\nodata&\nodata&\nodata&2\nl
721-10&135328.8&091436&\nodata&\nodata&\nodata&\nodata&\nodata&\nodata&2\nl
721-11&135404.2&084000&\nodata&\nodata&\nodata&\nodata&\nodata&\nodata&2\nl
721-12&135107.2&083554&6942&6908&112&D&861&9.27&2\nl
721-14&140213.2&090136&1196&1173&91&D&5716&8.55&2\nl
721-15&141042.0&085100&7393&7376&138&D&5590&10.14&2\nl
721-16&140038.4&082842&5260&5234&94&D:&1018&9.10&2\nl
721-17&135516.8&080454&\nodata&\nodata&\nodata&\nodata&\nodata&\nodata&2\nl
\nl
723-1&143121.6&130418&\nodata&\nodata&\nodata&\nodata&\nodata&\nodata&2\nl
723-2&144204.2&124236&\nodata&\nodata&\nodata&\nodata&\nodata&\nodata&2\nl
723-3&142857.0&122706&7919&7932&78&D:&1343&9.58&2\nl
723-4&143004.8&114618&2166&2177&109&D&3487&8.87&2\nl
723-5&143415.6&114736&1788&1803&100&D:&1051&8.19&2\nl
723-6&143255.8&084330&2092&2094&125&D&2473&8.69&2\nl
723-7&145003.6&102400&6922&6945&164&D&2622&9.75&2\nl
723-8&144806.6&083430&\nodata&\nodata&\nodata&\nodata&\nodata&\nodata&2\nl
723-9&144254.6&080254&1690&1774&33&S&4311&8.79&2\nl
\nl
749-2&230835.7&110540&2639&2853&69&S&1500&8.74&1\nl
749-3&231011.8&102729&6579&6790&129&D&1700&9.55&1\nl
749-4&231017.4&121322&4891&5108&142&D&1800&9.32&1\nl
749-6&231455.1&072139&3872&4071&91&D&3500&9.42&1\nl
749-7&231504.3&092254&3811&4016&127&D&1400&9.01&1\nl
749-8&231517.4&110905&3794&4005&68&S:&366&8.42&3\nl
749-9&231625.2&073021&4310&4508&102&D:&820&8.87&1\nl
749-10&231638.7&072550&3762&3960&159&D&3600&9.40&1\nl
749-11&231639.2&094449&\nodata&\nodata&\nodata&\nodata&\nodata&\nodata&3\nl
749-12&231941.4&090632&3574&3776&83&D&2700&9.24&1\nl
749-13&232005.4&100635&3521&3727&127&D&1700&9.03&1\nl
749-14&232015.5&112954&3836&4046&154&D&2000&9.17&1\nl
749-15&232639.9&113428&3648&3855&76&D:&1800&9.08&1\nl
749-17&233211.3&123909&3838&4046&119&D&1200&8.95&1\nl
\nl
774-1&074700.0&064418&4953&4793&162&D&2720&9.45&2\nl
774-2&074540.7&041342&5555&5385&137&D&1740&9.36&2\nl
774-3&073928.1&030524&7417&7245&156&D&2370&9.75&2\nl
\nl
822-4&233641.6&040800&6391&6567&197&D&1857&9.56&1\nl
822-6&233736.1&070153&3699&3885&92&D:&2600&9.25&1\nl
822-7&233840.1&053849&1732&1913&86&D:&1700&8.45&1\nl
822-9&234108.9&062402&3809&3991&138&D&1000&8.86&1\nl
822-10&234250.4&063315&5308&5490&84&D&1700&9.36&1\nl
822-11&234259.8&060813&4421&4597&178&D:&1554&9.17&3\nl
822-14&234646.8&035412&2971&3141&73&D:&570&8.40&1\nl
822-17&234811.6&034227&2907&3077&42&S&1362&8.76&3\nl
\nl
\enddata

\end{deluxetable}

\clearpage

\begin{deluxetable}{lccccc}

\tablecolumns{6}
\tablewidth{0pc}
\tablecaption{Optical and Mass Parameters}
\tablehead{
\colhead{Name} &
\colhead{Radius} &
\colhead{b/a} &
\colhead{$M_I$} &
\colhead{log $\Sigma_{HI}$} &
\colhead{log $M_{dyn}$} \\ 
\colhead{} &
\colhead{(arcsec)} &
\colhead{} &
\colhead{} &
\colhead{($M_{\sun}$ pc$^{-2}$}) &
\colhead{($M_{\sun}$}) \\
\nl
\colhead{(1)} &
\colhead{(2)} &
\colhead{(3)} &
\colhead{(4)} &
\colhead{(5)} &
\colhead{(6)} 
} 
\startdata
476-8 & 16.0&0.53& $-$13.9& 4.84& 8.05\nl
495-1 & 15.6&0.59& $-$16.4& 1.60& 9.21\nl
495-2 & 14.2&0.76& $-$16.3& 2.60& 8.60\nl
495-3 & 22.1&0.54& $-$15.6& 1.23& 8.98\nl
500-3 & 17.3&0.60& $-$16.3& 3.44& 9.22\nl
500-4 & 14.1&0.66& $-$16.4& 1.78& 8.93\nl
500-5 & 22.8&0.51& $-$19.0& 6.27&10.44\nl
508-2 & 25.8&0.76& $-$17.4& 3.68& 9.65\nl
508-5 & 16.8&0.74& $-$18.0& 2.36& 9.80\nl
512-2 & 25.4&0.50& $-$15.9& 4.38& 8.99\nl
512-3 & 12.0&0.48& $-$17.0& 6.62& 9.78\nl
512-7 & 12.8&0.82& $-$17.7& 10.93&10.23\nl
512-9 & 17.7&0.73& $-$18.5& 4.71&10.18\nl
512-10& 11.6&0.61& $-$17.0& 3.84& 9.43\nl
563-1 & 14.7&0.83& $-$17.4& 1.76& 8.92\nl
563-4 & 26.2&0.57& $-$18.1& 7.76& 9.99\nl
563-5 & 18.1&0.37& $-$18.3& 13.32& 9.86\nl
564-4 & 14.6&0.62& $-$16.8& 2.52& 9.65\nl
564-8 & 15.1&0.83& $-$13.4& 5.75& 8.41\nl
564-11&  7.3&0.57& $-$15.7& 18.62& 9.85\nl
564-12& 16.9&0.57& $-$18.1& 8.30&10.11\nl
564-15& 18.1&0.53& $-$17.6& 3.84& 9.99\nl
565-1 & 11.5&0.36& $-$18.3& 13.71& 9.73\nl
565-10& 22.0&0.85& $-$13.8& 1.07& 8.13\nl
568-2 & 13.1&0.64& $-$19.0& 3.21& 8.25\nl
568-4 & 24.4&0.85& $-$17.1& 1.52& 9.54\nl
570-4 & 24.8&0.68& $-$15.5& 2.65& 8.99\nl
572-2 &  9.9&0.23& $-$16.2& 19.01& 9.73\nl
572-4 & 13.5&0.80& $-$15.0& 3.57& 8.44\nl
572-5 & 17.4&0.66& $-$15.3& 6.47& 8.79\nl
575-1 & 26.6&0.51& $-$13.6& 3.91& 8.25\nl
575-2 & 40.1&0.49& $-$16.6& 6.24& 9.52\nl
575-5 & 24.6&0.44& $-$12.0& 9.79& 8.22\nl
575-7 & 17.3&0.57& $-$15.2& 8.50& 8.89\nl
576-9 & 13.4&0.64& $-$16.6& 6.60& 9.80\nl
577-2 & 12.6&0.64& $-$18.2& 8.18& 9.86\nl
577-5 & 15.7&0.61& $-$17.6& 6.04& 9.88\nl
582-1 &  8.4&0.36& $-$16.9& 54.44&10.61\nl
582-2 &  8.1&0.87& $-$16.5& 2.87& 9.16\nl
584-2 & 21.3&0.62& $-$17.5& 4.33& 9.84\nl
584-3 & 17.3&0.49& $-$16.6& 7.10& 9.80\nl
584-4 & 23.8&0.38& $-$17.5& 21.55&10.11\nl
631-1 & 18.5&0.69& $-$18.5& 4.05&10.29\nl
631-7 & 42.5&0.44& $-$14.4& 11.54& 8.66\nl
631-8 & 11.5&0.64& $-$17.0& 6.59& 9.94\nl
637-18& 21.0&0.82& $-$16.9& 2.41& 9.04\nl
640-7 & 15.5&0.47& $-$15.5& 19.85& 9.09\nl
640-13& 21.1&0.68& $-$14.5& 2.40& 8.48\nl
646-2 & 13.8&0.67& $-$17.5& 15.32&10.02\nl
646-7 & 65.3&0.59& $-$12.2& 1.67& 8.11\nl
646-8 &  9.2&0.48& $-$15.0& 1.71& 8.55\nl
646-11& 25.3&0.85& $-$13.0& 2.20& 8.80\nl
656-2 & 31.0&0.52& $-$16.0& 4.89& 9.49\nl
702-1 & 18.5&0.89& $-$16.3& 5.38& 9.73\nl
704-3 & 15.4&0.58& $-$17.3& 5.08& 9.47\nl
709-10& 14.1&0.68& $-$16.0& 1.86& 8.64\nl
709-11&  9.3&0.47& $-$15.2& 3.26& 8.83\nl
721-5 & 13.8&0.69& $-$17.7& 4.33& 9.89\nl
721-16& 15.3&0.71& $-$18.2& 3.66& 9.91\nl
723-5 & 27.0&0.58& $-$17.5& 1.49& 9.64\nl
723-6 & 17.4&0.36& $-$16.3& 13.55& 9.63\nl
723-9 & 36.1&0.44& $-$17.6& 4.52& 8.88\nl
774-2 & 15.6&0.39& $-$18.3& 11.01&10.09\nl
774-3 & 17.2&0.75& $-$18.6& 6.38&10.67\nl
\nl
\enddata

\end{deluxetable}

\clearpage

\clearpage 
\begin{figure} 
\plotfiddle{profiles.1}{11.5truein}{0}{100}{100}{-310}{170} 
\end{figure} 

\clearpage 
\begin{figure} 
\plotfiddle{profiles.2}{11.5truein}{0}{100}{100}{-310}{170} 
\end{figure} 

\clearpage 
\begin{figure} 
\plotfiddle{profiles.3}{11.5truein}{0}{100}{100}{-310}{170} 
\end{figure} 

\clearpage 
\begin{figure} 
\plotfiddle{profiles.4}{11.5truein}{0}{100}{100}{-310}{170} 
\end{figure} 

\clearpage 
\begin{figure} 
\plotfiddle{profiles.5}{11.5truein}{0}{100}{100}{-310}{170} 
\end{figure} 

\clearpage 
\begin{figure} 
\plotfiddle{profiles.6}{11.5truein}{0}{100}{100}{-310}{170} 
\end{figure} 

\clearpage 
\begin{figure} 
\plotfiddle{profiles.7}{11.5truein}{0}{100}{100}{-310}{170} 
\end{figure} 

\clearpage 
\begin{figure} 
\plotfiddle{profiles.8}{11.5truein}{0}{100}{100}{-310}{170} 
\end{figure} 

\clearpage 
\begin{figure} 
\plotfiddle{profiles.9}{11.5truein}{0}{100}{100}{-310}{170} 
\end{figure} 

\clearpage 
\begin{figure} 
\plotfiddle{profiles.10}{11.5truein}{0}{100}{100}{-310}{170} 
\end{figure} 

\clearpage 
\begin{figure} 
\plotfiddle{profiles.11}{11.5truein}{0}{100}{100}{-310}{170} 
\end{figure} 

\clearpage 
\begin{figure} 
\plotfiddle{profiles.12}{11.5truein}{0}{100}{100}{-310}{170} 
\end{figure} 

\clearpage 
\begin{figure} 
\plotfiddle{profiles.13}{11.5truein}{0}{100}{100}{-310}{170} 
\end{figure} 

\clearpage 
\begin{figure} 
\plotfiddle{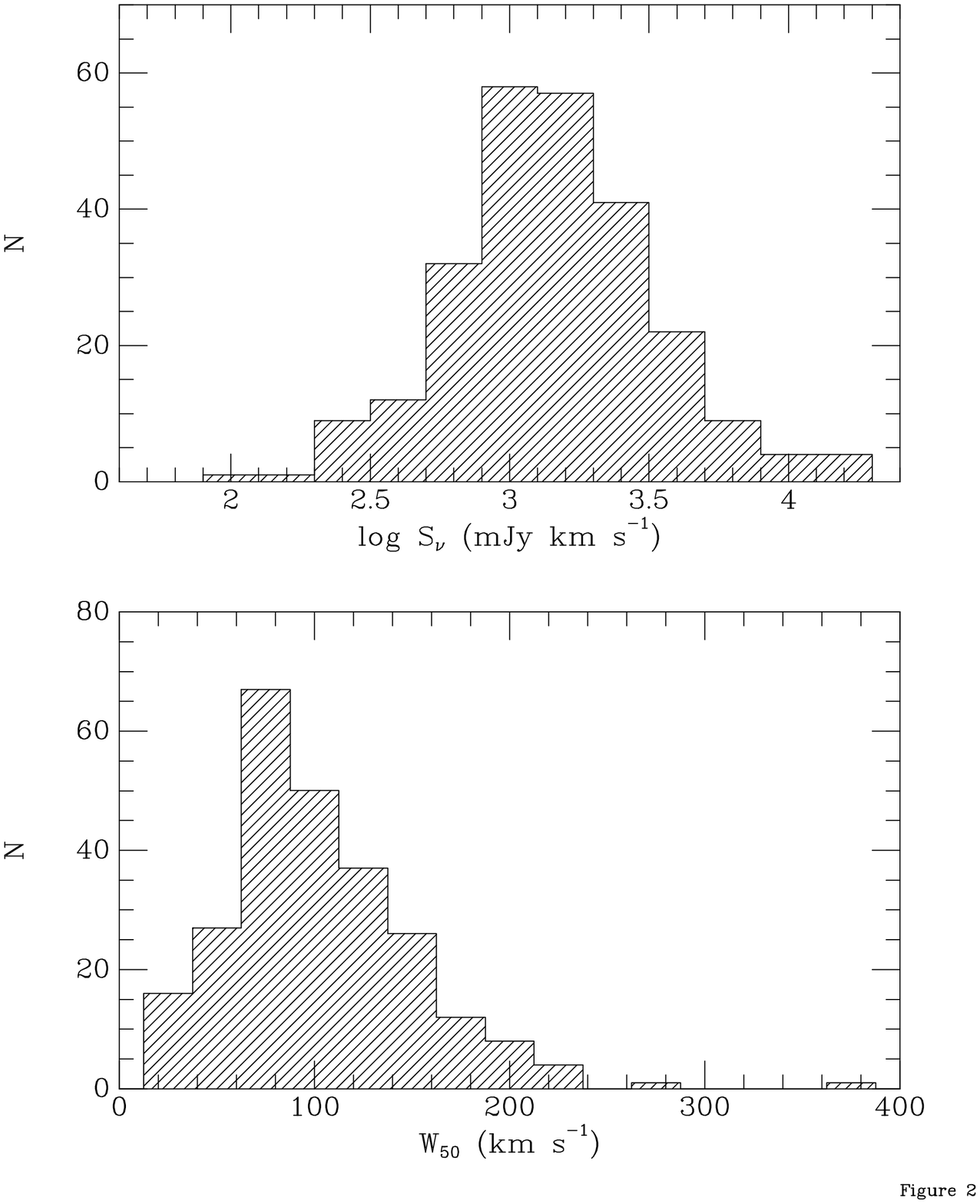}{11.5truein}{0}{100}{100}{-310}{170} 
\end{figure} 

\clearpage 
\begin{figure} 
\plotfiddle{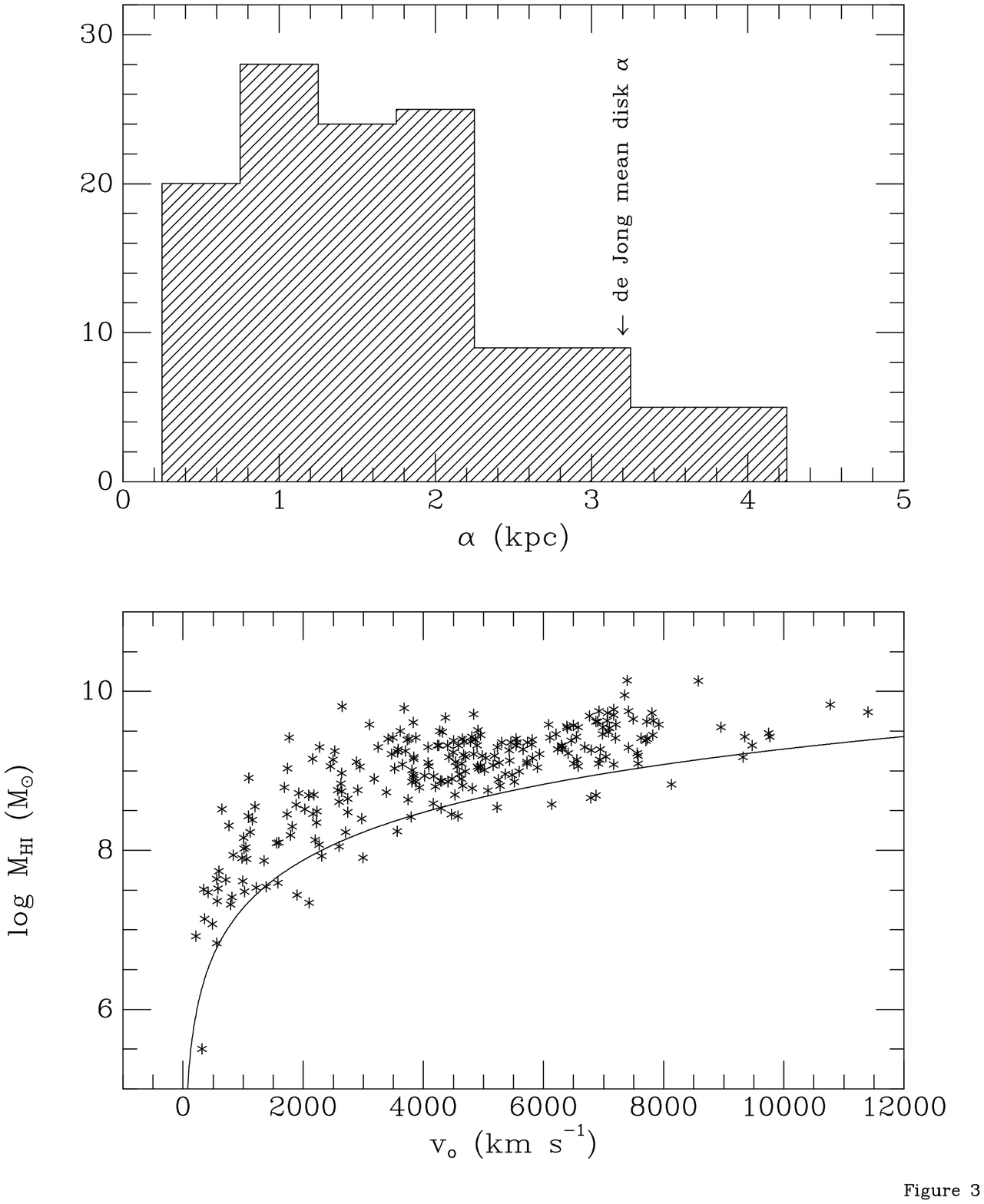}{11.5truein}{0}{100}{100}{-310}{170} 
\end{figure} 

\clearpage 
\begin{figure} 
\plotfiddle{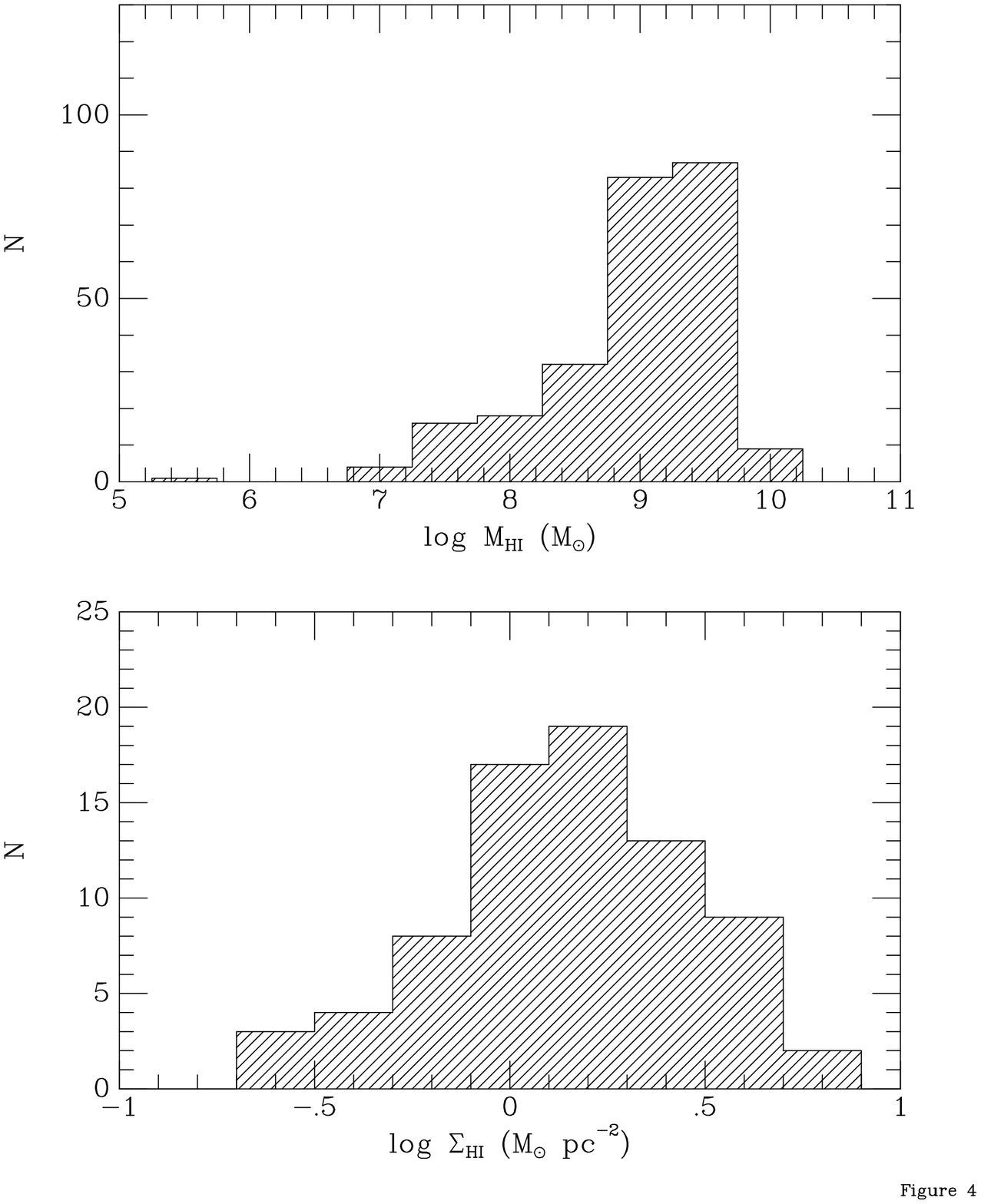}{11.5truein}{0}{100}{100}{-310}{170} 
\end{figure} 

\clearpage 
\begin{figure} 
\plotfiddle{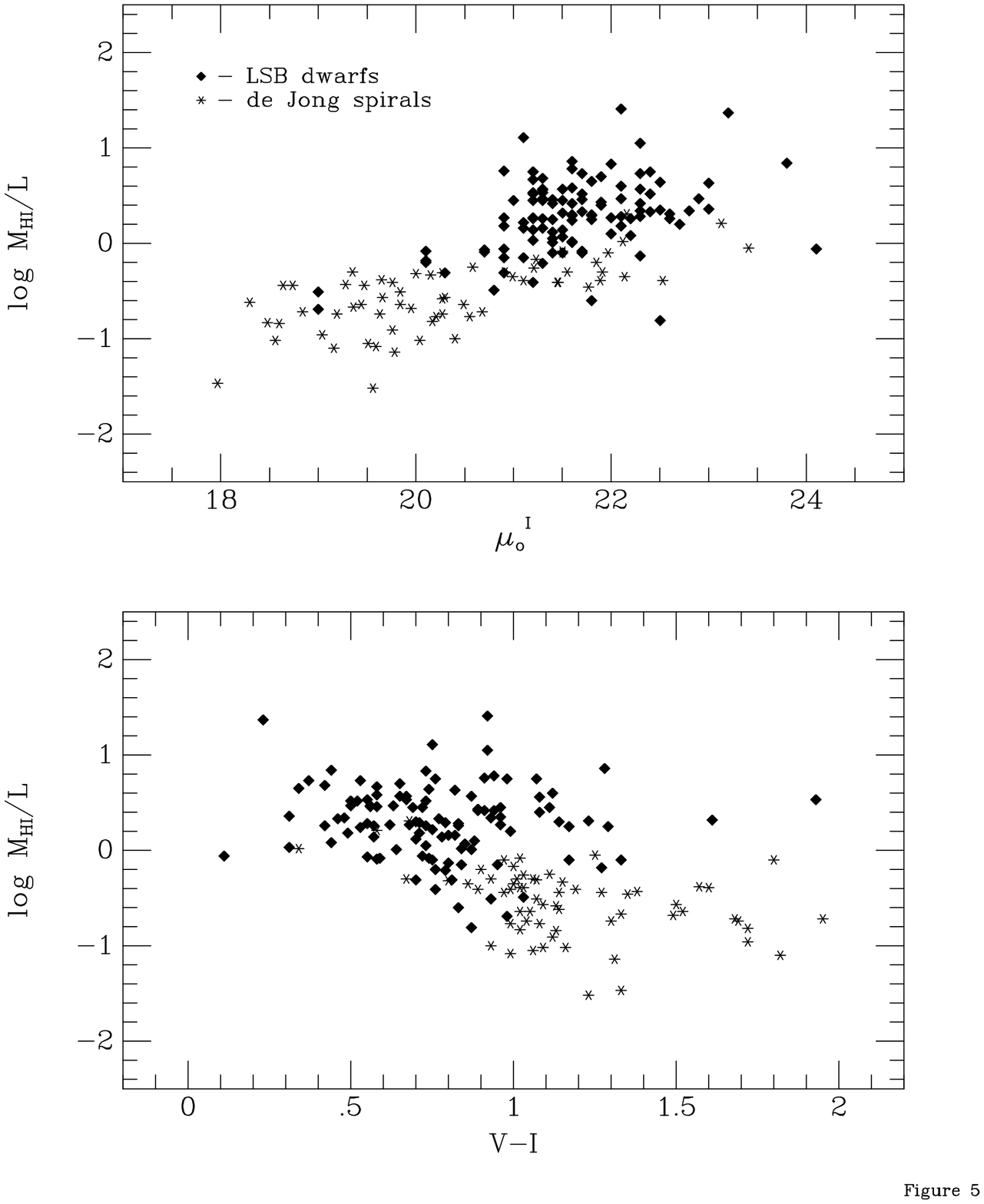}{11.5truein}{0}{100}{100}{-310}{170} 
\end{figure} 

\clearpage 
\begin{figure} 
\plotfiddle{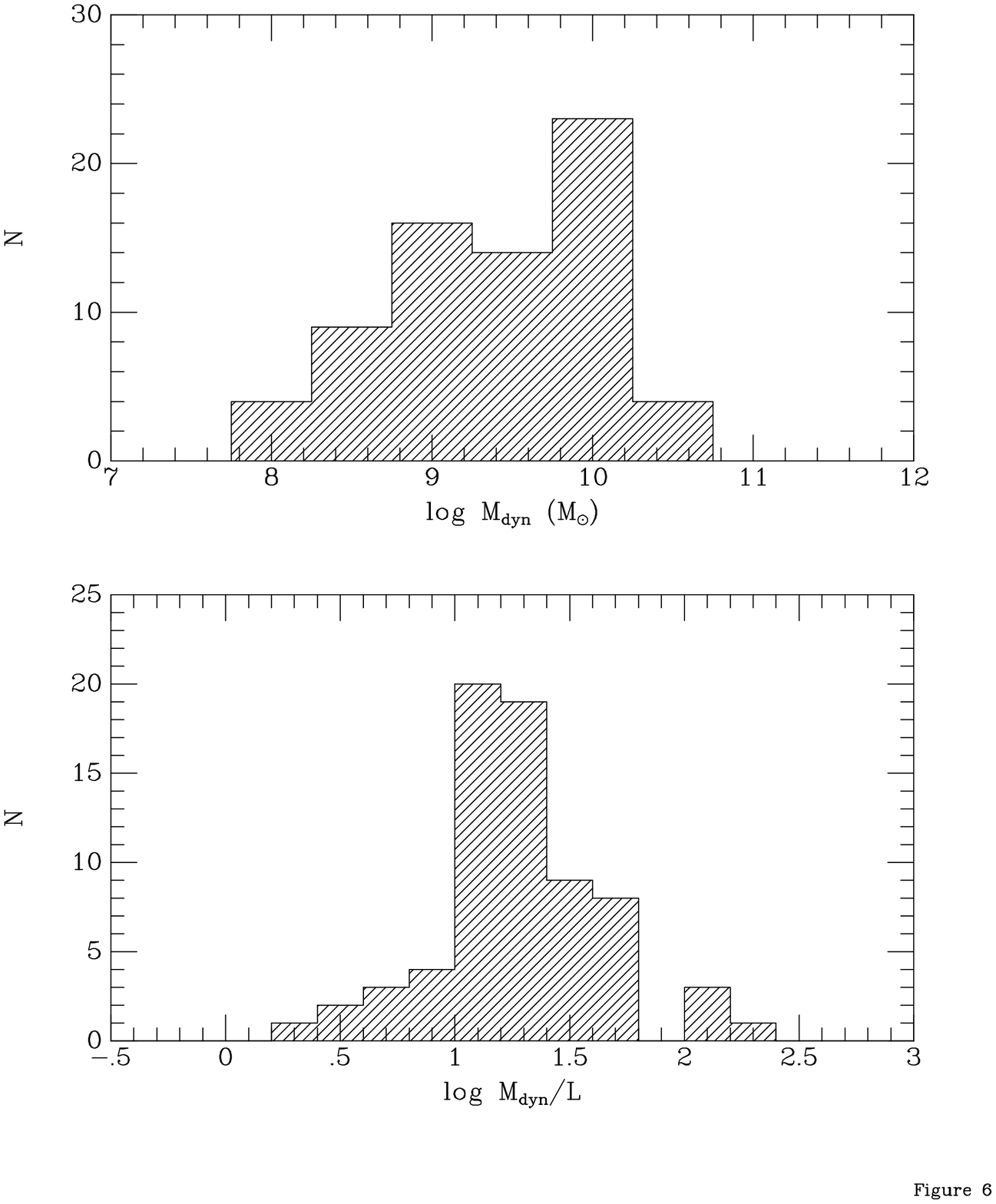}{11.5truein}{0}{100}{100}{-310}{170} 
\end{figure} 

\clearpage 
\begin{figure} 
\plotfiddle{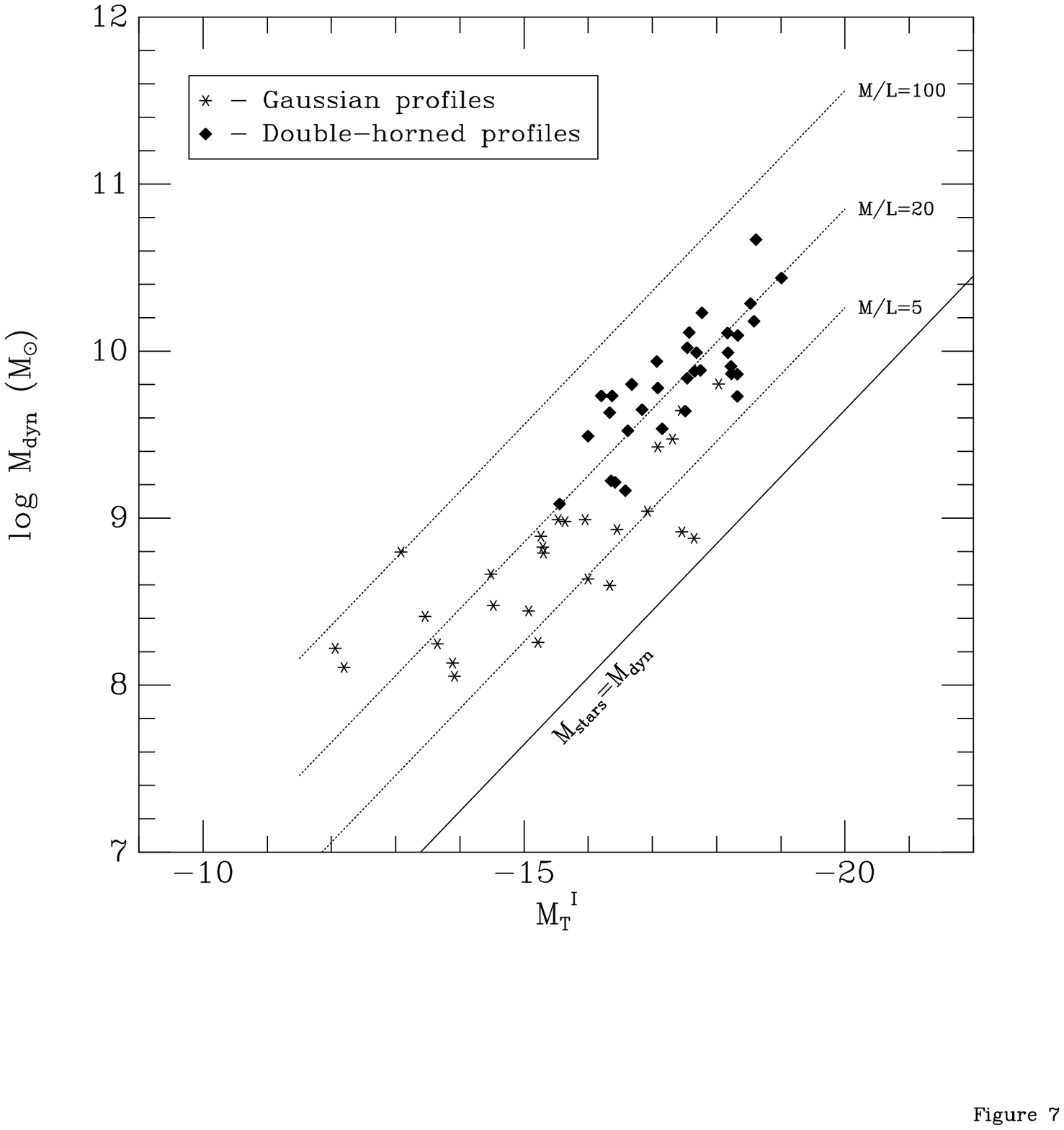}{11.5truein}{0}{100}{100}{-310}{170} 
\end{figure} 

\clearpage 
\begin{figure} 
\plotfiddle{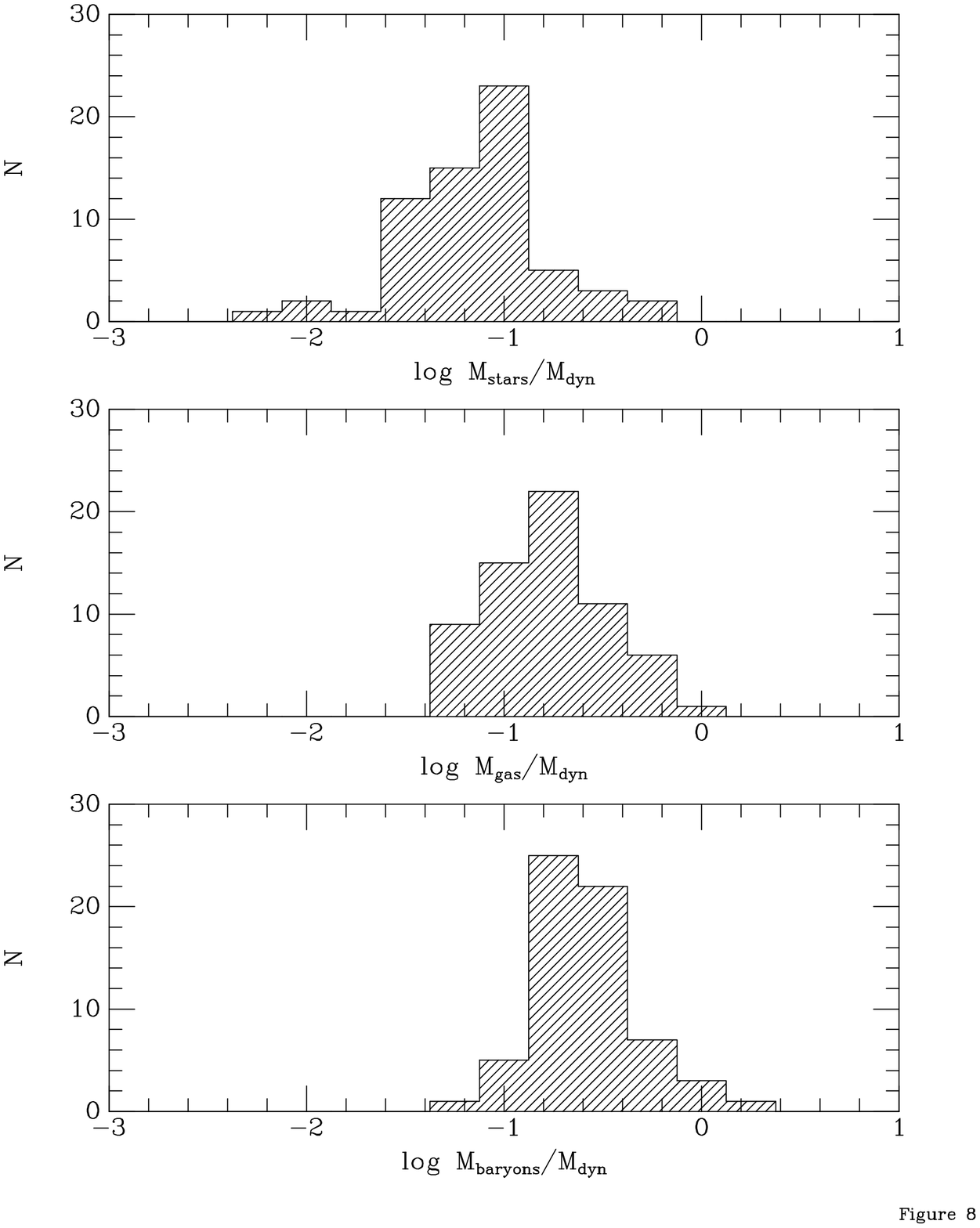}{11.5truein}{0}{100}{100}{-310}{170} 
\end{figure} 

\end{document}